\documentclass[acmlarge,screen]{acmart}

\AtBeginDocument{%
  \providecommand\BibTeX{{%
    \normalfont B\kern-0.5em{\scshape i\kern-0.25em b}\kern-0.8em\TeX}}}

\begin{document}


\def \projectName {{CasualGaze}}
\title{\projectName{}: Towards Modeling and Recognizing Casual Gaze Behavior for Efficient Gaze-based Object Selection}


\author{Yingtian Shi}
\authornote{Both authors contributed equally to this research.}

\author{Yukang Yan}
\authornotemark[1]
\affiliation{%
  \institution{Tsinghua University}
  \country{China}
}

\author{Zisu Li}
\affiliation{%
  \institution{Tsinghua University}
  \country{China}
}

\author{Chen Liang}
\affiliation{%
  \institution{ Tsinghua University}
  \country{China}
}
\author{Yuntao Wang}
\affiliation{%
  \institution{Tsinghua University}
  \country{China}
}
\author{Chun Yu}
\authornote{Corresponding author.}
\affiliation{%
  \institution{Tsinghua University}
  \country{China}
}

\author{Yuanchun Shi}
\affiliation{%
  \institution{Tsinghua University}
  \country{China}
}



\begin{abstract}
We present CasualGaze, a novel eye-gaze-based target selection technique to support natural and casual eye-gaze input. Unlike existing solutions that require users to keep the eye-gaze center on the target actively, CasualGaze allows users to glance at the target object to complete the selection simply. To understand casual gaze behavior, we studied the spatial distribution of casual gaze for different layouts and user behavior in a simulated real-world environment. Results revealed the impacts of object parameters, the speed and randomness features of casual gaze, and special gaze behavior patterns in "blurred areas". Based on the results, we devised CasualGaze algorithms, employing a bivariate Gaussian distribution model along with temporal compensation and voting algorithms for robust target prediction. The usability evaluation study showed significant improvements in recognition and selection speed for CasualGaze compared with the two baseline techniques. Subjective ratings and comments further supported the preference for CasualGaze regarding efficiency, accuracy, and stability.

\end{abstract}


\begin{teaserfigure}
  \includegraphics[width=\textwidth]{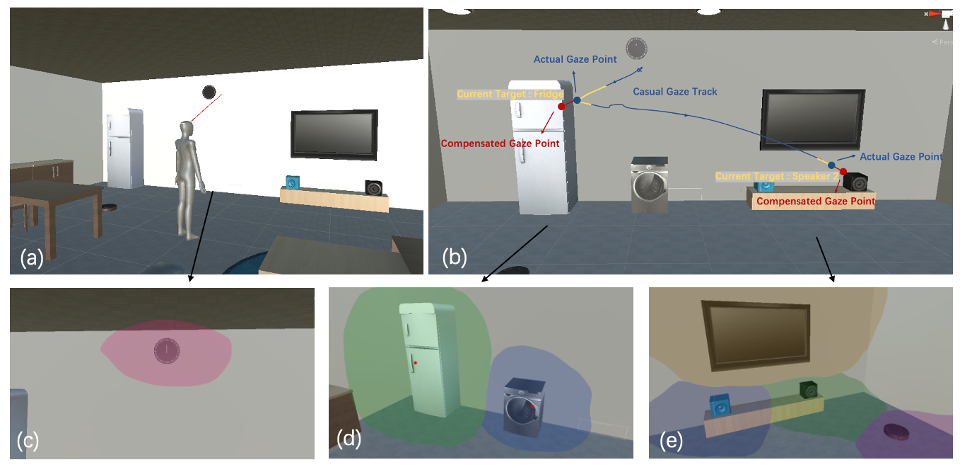}
  \caption{We propose \projectName{}, a behavior model of user's casual gaze interaction, which predicts user intent through the distribution of the user's casual gaze when interacting with the device in the smart environment(a). We give an example of interaction with \projectName{} (b), which identifies the interaction target in advance by predicting the user's gaze in conjunction with the distribution around devices. We establish a two-dimensional Gaussian model for gaze, which will be adjusted according to the device's location, size, and surrounding device information. It is suitable for isolated devices (c), a pair of devices (d), and multiple devices (e).}
  \label{fig:teaser}
\end{teaserfigure}

\maketitle


\section{Introduction}

The emergence of heterogeneous smart devices in our surrounding environment casts the demand for general access techniques to facilitate natural and efficient interaction. Among the possible input modalities, gaze is considered one with great potential to enable efficient interaction in VR, AR, and IoT scenarios \cite{majaranta2019eye, barbuceanu2011attentive}. 

Gaze-based interaction has been widely researched in previous literature \cite{looktotalk, gaze-voiceincar, EyePliances, MediaEyePliances}, showing the efficacy of using gaze as an always-available and intuitive input channel in target selection \cite{mayer2020enhancing}, region-of-interest indication \cite{10.1145/3544548.3580696}, and gesture expression \cite{yi2022gazedock}. Among the vast gaze-related research topics, target selection with eye gaze is the most challenging one due to the eye gaze's uncertainty and the Midas problem of detecting gaze intention from continuous gaze points \cite{hansen2009eye}.
Existing solutions of gaze-based techniques usually require the user to indicate their attention explicitly by keeping their eye gaze point within a certain distance from the target, which conflicts with eye gaze behavior nature and may bring extra mental burden.

In this work, we present CasualGaze, a novel eye gaze-based target selection technique to support natural and casual eye-gaze input with lower attention levels and mental load. With CasualGaze, when the user wants to select an object in the scene, they merely need to move their gaze toward or have glance at the object in a casual manner. During the gaze point movement, the system reports the recognition (e.g., by highlighting the target object) at an earlier time point (e.g., when glimpsing at the object while the gaze is still approaching it) so that the user could confirm their intention at the earlier stage and perform subsequent interactions, as shown in Figure \ref{fig:teaser}.

To understand casual gaze behavior, we first conducted a user study to investigate end-point characteristics of casual gaze behavior for independent objects as well as the interacting effect of two nearby objects. Results found that relative location, distance, and size had impacts on either center shift or spread size of gaze distribution. We further conducted a second study to collect and analyze casual gaze behavior data in a room-scale environment with 18 target devices, where we derived speed and randomness features and found special behavior patterns in "blurred areas" for casual gaze. 

Based on casual gaze modeling results, we devised a target selection algorithm consisting of a bivariate Gaussian distribution model to model the spreads of objects with different sizes and relative locations, as well as a temporal compensation algorithm and a voting algorithm for robust target prediction. Results from offline simulation showed that \projectName{} algorithms achieved an average accuracy of 96.47\% in target prediction, which was significantly higher than the KNN baseline and close to the optimal model built from the prior gaze point distributions with the entire dataset.

Finally, we evaluated the real-time performance and user experience of \projectName{} target selection compared with two baseline techniques - precise selection and KNN-based selection - in three real-life scenarios with different target layouts (home, classroom, and office) throughout a usability evaluation study. Results found \projectName{} significantly outperformed two baselines in both recognition speed (e.g., 1.08s V.S. 1.35s V.S. 1.73s) and selection speed (e.g., 1.65s V.S. 1.77s V.S. 1.84s). Subjective ratings and comments showed the user's preference of \projectName{} in aspects of efficiency, accuracy, and stability. 

To sum up, we figure out three main contributions in this paper:
\begin{enumerate}
    \item We are the first to present the concept of casual gaze, which we think is an important concept to facilitate natural gaze-based interaction. Through two user studies, We modeled users' casual gaze behavior both in spatial and temporal domains. 
    \item We proposed novel algorithms to implement \projectName{}, which integrates the Gaussian model into the Bayesian inference method to predict the target based on casual gaze data.
    \item We evaluate the performance of \projectName{} through both offline simulation and real-time device selection tasks. Results show that \projectName{} improves the device selection accuracy, speed, and interaction experience, and it's extendable and applicable to different device spatial layouts.
\end{enumerate}

\section{Related Work}

\subsection{Pointing and Target Selection in IoT Scenarios}

Smart devices are becoming increasingly available to users, and connected smart devices make up a smart environment that facilitates the users to gain information and take actions\cite{whydevice}. Related research explored techniques with different modalities to interact with smart devices. Remote control \cite{remotecontrol} is a typical device specially designed for smart device control. Smartphones\cite {phoneapp, arhome}, tablet\cite{phonetablet}, and hand-held wand\cite{xwand} are replacing remote controls as they can provide visual or haptic feedback during control. Luria et al. \cite{socialrobot} reported social robots can provide an engaging device control interface with high situation awareness. Without a specific control device, voice input \cite{voicehome1, voicehome2} is popular in the interaction between users and smart devices, which can also be used in combination with hand gestures \cite{gesturevoice, gesturehome}. Voice is preferred by users as it is intuitive and direct, and users can speak voice commands to control the devices. However, it is inconvenient to specify the device to interact only via voice commands and lack of support for device disambiguation can cause false positive responses from multiple smart devices\cite{voicedirection}. The selection and interaction of smart devices based on user gaze have also been widely studied. Users can select devices more directly through the direction of gaze \cite{AmbiGaze} and control the devices through gestures \cite{watchdo, isomoto2020gaze}. Jungwirth et al. \cite{jungwirth2018eyes} illustrate that gaze is a better long-distance interaction method than touch, but accurate eye movement depends on the smoothing method and calibration. We believe that there are specific rules in the user's gaze behavior, and efficient target selection technology can be achieved without relying on accurate eye movement data. \projectName{} provides a more natural gaze interaction scheme by analyzing the user's gaze data. Since the user is not required to look at the target accurately, it also reduces the cost of smoothing and calibration.


\subsection{Understanding Interactive Gaze Behavior}



Gaze is an indispensable input modality to facilitate convenient and natural interaction with distributed objects and devices in VR, AR, and IoT scenarios \cite{majaranta2019eye, barbuceanu2011attentive, eyegazeadv, gazebasedpre, hgazetyping, eyespyvr}. Users can complete accurate gaze control based on the VestibuloOcular Reflex (VOR)\cite{leigh2015neurology}. Unfortunately, the current use of gaze for interaction in commodity devices such as Microsoft Hololens \cite{van2017we, blattgerste2018advantages} and Oculus VR \cite{atienza2016interaction} is largely restricted to simple logic defined by the present determined gaze point regardless of the user's gaze behavior. To fill this gap, previous work has studied general or application-specific gaze behavior in different dimensions, including pointing accuracy \cite{kumar2008improving} and gaze point stability \cite{koh2009input, CircularOrbits}.


Head-gaze \cite{Zhang14HOBS, headcross, WorldGaze,realtimehead, headgazeheadges}and eye-gaze \cite{Zhai99Gaze,10.1145/332040.332445, dwellmlgaze} can provide a natural and convenient addition to the selection of an object of interest in user-computer dialogue. Yan et al. \cite{headcross} uses the cross movement of the head-gaze at the boundary of the target to complete the selection. With the help of head-gaze, Mayer et al. \cite{WorldGaze} uses mobile devices to capture the user's object of interest. Eye-gaze is better than head-gaze in terms of speed and task load\cite{blattgerste2018advantages, 10.1145/332040.332445}. Zhai et al. \cite{Zhai99Gaze} combine eye movement with manual to achieve more precise computer input.
Gaze motion is achieved through the seamless coordination of the eyes and head \cite{headeyetorso, eyehead}. Furthermore, the combination of head-gaze and eye-gaze enables the more sophisticated interaction\cite{BimodalGaze, radieye}. Sidenmark and Gellersen \cite{headeyetorso} propose the analysis of the contribution and time alignment of the relative movement of the eyes, head, and torso in VR, which is consistent with previous studies in actual scenes \cite{tomlinson1986combined, leigh2015neurology}. Pinpointing \cite{pinpointing} proposes a comparison of speed and pointing accuracy, revealing the relative advantages of multimodal selection based on head and eye movements, including a robust selection of achievable target sizes. Chang and Hu's \cite{chang2003multivariable} work uses the eye-head torso target tracking system to demonstrate how to achieve cooperative control and fault adjustment. These technologies are all based on the smooth and calibrated precise gaze orientation.

To our knowledge, the behavioral characteristics (e.g., the gaze trajectory and the gaze point distribution) of how the user casually glances at the target object in different spatial contexts with interactive intention have never been researched. In this work, we systematically conducted four user studies to build models for users' gaze behavior in different target spatial layouts and provide instructions for user intention judgments. Our study results and insights could serve as a complement to existing works.

\subsection{Attentive User Interface}
In human-computer interaction, a line of research focuses on attentive user interfaces that adaptively adjust the interfaces according to different states of user attention \cite{selker2004visual, looktotalk, gaze-voiceincar, EyePliances, MediaEyePliances}. They enable users to start conversations with software agents\cite{looktotalk}, select applications on computers\cite{MediaEyePliances, gaze-voiceincar}, control home appliances \cite{EyePliances,auralamp} by looking to the targets. In the choices of expressing the intention, eye gaze is frequently used to express the intention to communicate\cite{gaze-voiceincar}, and is the most reliable indication of the target of interest \cite{eyepattern}. A framework \cite{AUIframework} is proposed to guide the design of attentive user interfaces. We regard \projectName{} as fundamental research that will facilitate these attentive user interfaces. As staring at the device is not the most intuitive way of expressing interaction intention, \projectName{} aims to loosen the requirements for users to trigger smart devices. With a statistical model, we determine in what range around the device that the user is looking suggests a strong intention to interact. Once the intention is recognized, the attentive interfaces can react adaptively.

\section{Study 1: Understanding Casual Gaze's Spatial Distribution}

In this section, we first introduce the concept of casual gaze. Then we conducted a two-part user study to investigate the spatial distribution of casual gaze behavior for independent objects and interacting object pairs. Results from this section gain insights into how the target parameters and spatial layouts influenced the distribution of casual gaze behavior.

\subsection{Definition of Casual Gaze}

The fundamental goal of GasualGaze is to allow the user to select and interact with objects distributed in the scene by casually looking toward the target instead of precisely looking at it. The casualness in our concept is reflected in two aspects. First, the user does not need to stare at, or actively keep the center of vision at the target object. Instead, they can simply perform a glance, either unconcentrated or with their peripheral vision, as their intuitive intention of reaching a specific target. Second, the user does not expect a "waiting and confirmation stage", meaning the system could potentially recognize the gaze intention before the gaze reaches the target. The user can fluently proceed to the subsequent interactions without waiting or needing to confirm while their gaze is in motion.

Specifically, we identify three important principles for GasualGaze:

1. Precise gaze behavior should be included and is a corner case of CasualGaze. Casual gaze refers to a more general and relaxed form of eye movement or gaze direction while precise gaze behavior represents a specific scenario where a higher level of accuracy and intentionality is required. Therefore, GasualGaze algorithms should be self-contained and capable of recognizing precise gaze behavior. 

2. CasualGaze models the gaze behavior as a process of temporal gaze point sequence. Different from traditional gaze modeling where the emphasis is often placed on predicting or analyzing a single gaze point, CasualGaze takes a different approach by considering the entire sequence of gaze points over time. By observing the gaze point sequence, CasualGaze can capture temporal features to solidify and accelerate the recognition. 

3. Casual gaze is an interacting behavior influenced by the quality (e.g., accuracy and stability) of the recognition system's real-time feedback. In real practice, how casually the user tends to perform their gaze selection is largely confined by how they perceive the system's recognition capability of their gaze intention, specifically as a way of feedback, which is an adaptation process. 

In the following subsections, we conducted a two-part study to investigate users' casual gaze behavior on 1) selecting individual objects with different shapes and at different locations and 2) selecting two nearby objects with interacting effects. 



\label{sec:stu_1}
\subsection{Part 1: Casual Gaze Behavior for Single Object}

The first part of the study investigated the user's gaze behavior for a single object in the scene with regard to different sizes and different relative locations.

\subsubsection{Participants and Apparatus}
We recruited 12 participants (7 females) with an average range of 25.41 (SD = 4.05) from the local campus via an online questionnaire. Eight participants had prior experience with head-mounted VR displays. 
We conducted our experiment with an HTC Vive Pro Eye, which features an eye tracker with a high tracking accuracy of 0.5 to 1.1 degrees. We developed our experiment platform with Unity 2019. 

\subsubsection{Design}
The experiment took part in a $7 m \times 7 m \times 4 m$ empty room in VR, with a sphere as the target appearing at different locations in the room. The basic task of the participant is to reach the displayed sphere with their eye gaze. The participant should press the controller's trigger as soon as they regarded themselves as having spotted the target to confirm target selection. Such a process yielded casual gaze data that satisfied the abovementioned casual gaze criteria. 


We adopted a within-subject factorial design to investigate the influence of target size and location on casual gaze distribution.
To cover the size range of common indoor objects such as smart IoT devices, we designed four levels - 0.25, 0.50, 0.75, and 1.00 meters for the target size. These sizes corresponded to common indoor devices from a smart speaker (0.25 meters in height) to an air conditioner (1.00 meters in width).
We sampled 24 locations ($8~horizontal~angles~\times~3~vertical~angles$) in the room to locate the target. We selected three levels in the vertical 
: 30 degrees down (e.g., cleaning robot on the floor), horizontal to the eye (e.g., TVs and smart speakers on tabletops), and 30 degrees up (e.g., light on the ceiling). We chose eight horizontal directions for each vertical angle separated by 45 degrees to cover all possible device locations.

Changing the distance between the target and the user causes the same result in the user's sight as changing the target size, so we fixed the distance from the target to 3 meters.
To help participants become aware of the target's location at the start, we also included a visual aid in the form of an arrow pointing toward the target. 

 \begin{figure}[htbp]
\centering
\begin{minipage}[t]{0.48\textwidth}
\centering
\includegraphics[width=6cm]{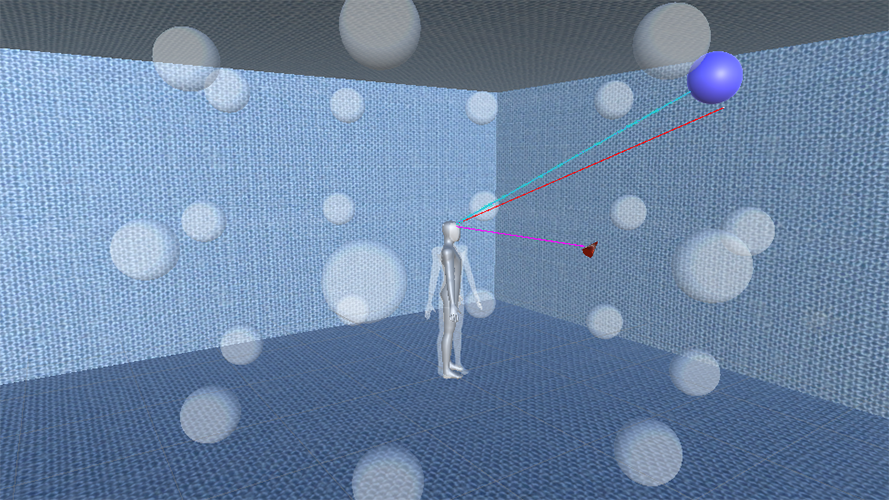}
\label{fig_22}

\end{minipage}
\begin{minipage}[t]{0.48\textwidth}
\centering
\includegraphics[width=6cm]{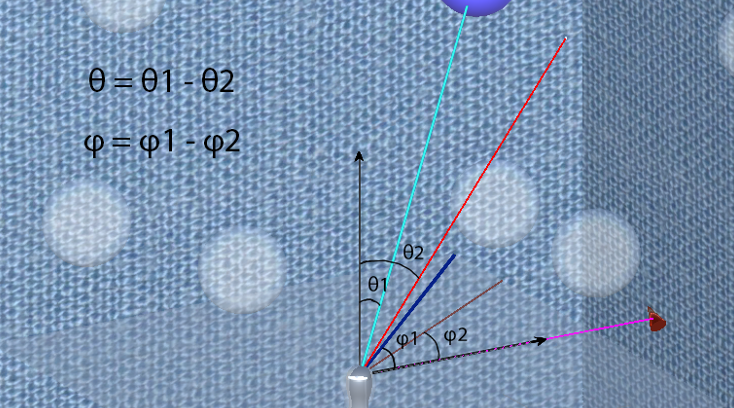}
\label{fig_22}

\end{minipage}

\caption{The target location set for Study 1 (left). Three heights and eight horizontal angles are set with the user as the center. The red cone indicates the target direction in a single trail. 
The offset of the user's gaze relative to the target direction(right). Calculate the horizontal ($\phi$) and vertical ($\theta$) angles of the user's gaze (red line) and the target direction (blue line). The difference between these two angles represents the offset of the user's gaze relative to the target.}
\end{figure}

\subsubsection{Task and Procedure}


An experimenter first introduces the experiment to the participants.
The participants wore a headset and calibrated the gaze tracker with the experimenter's help. A virtual room centered on the user's head appeared when they looked straight ahead and pulled the trigger. The participants were given five minutes to familiarize themselves with the virtual room and the interaction for selection. 

At the start of each trial, the participants moved to the starting position and looked toward the starting direction. After a three-second countdown, an arrow indicating the target position appeared, and the participants began performing the task. Once they pressed the button to confirm their selection, they reset their gaze direction, and a new trial started. The participants were instructed to look at the target in a natural and casual manner while pressing the trigger as quickly as possible when they felt they had spotted the target. 

Each participant performed $10~sessions~\times4~target~sizes~\times~24~target~positions~=~960~tasks$.
The order of the target sizes and positions were randomized in each session for each participant. 
The whole study took around 50 minutes, with a break given every two sessions. Participants' gaze data, as well as the trigger timestamps were recorded during the whole study.
Each participant received a 16 USD gift card for compensation.

\subsubsection{Results}
We obtained a total of 5605 selection data samples in the experiment and removed user-reported errors and large deviations in the overall statistics (107 times, 1.91\%).
To analyze the distribution of the user's gaze around the target, we calibrated the results and found that the angle offset in the horizontal and vertical directions followed a two-dimensional Gaussian distribution. The covariance of the two directions is close to 0, so we mainly consider the mean and standard deviation in both directions. We ran RM-ANOVA tests ($p < 0.05$) on the metric of angular offsets $(\Phi{s}, \Theta{s})$ with the variables of target size and target location, with the post-hoc T-tests ($p < 0.05$).
We highlight our significant results and findings below.

\textbf{1) Target size has no significant influence on the distribution.} 
RM-ANOVA tests showed that the size of the target has no significant effect on both the mean and variance in the two directions ($\Theta{mean}:F_{(3,33)} = 0.723, ~p =0.546, \Theta{std}:F_{(3,33)} = 0.284, ~p =0.836, \Phi{mean}:F_{(3,33)} = 0.723, ~p =0.546, \Phi{std}:F_{(3,33)} = 1.561, ~p =0.217$). When only one device is in the space, its size will not cause a significant change in the user's casual gaze behavior. 
If the device is large, it is easily noticeable to the user, while a small device may not be accurately seen by the user. Users tend to assume they can express their intention to select the device without accurately seeing its inside. Ten users in the post-study interview confirmed this, who reported that they could express their intention even without staring at it accurately.


\textbf{2) Target position significantly affects the angular offset in a symmetrical manner.} 
RM-ANOVA tests showed that the target's height and horizontal angle significantly affect the offset's overall mean and standard deviation ($\text{Height}: \text{Offset}_{mean}:F_{(2,22)} = 12.668, ~p < 0.05, \text{Offset}_{std}:F_{(2,22)} = 12.626, ~p <0.05, \text{Horizontal Angle}: \text{Offset}_{mean}:F_{(7,77)} = 7.355, ~p <0.05, \text{Offset}_{std}:F_{(7,77)} = 10.167, ~p =0.<0.05,$). 

The height only significantly affects the mean and variance of the offset in the vertical direction ($\Theta{mean}:F_{(2,22)} = 5.986, ~p <0.05, \Theta{std}:F_{(2,22)} = 26.387, ~p <0.05, \Phi{mean}:F_{(2,22)} = 5.986, ~p = 0.090, \Phi{std}:F_{(2,22)} = 6.199, ~p = 0.074$). 
We observed that the center of the user's gaze distribution shifted towards the initial gaze direction, indicating that users prefer less body or eye movement to express intention. This aligns with our original intention to study users' casual gaze behavior. Specifically, when the target is higher, the user's gaze tends to lean toward the lower part of the target. Changes in height did not produce a significant deviation in the horizontal angle of the user's gaze. This can be attributed to the fact that the three height levels were all within the user's comfortable eye movement range, resulting in consistent gaze behavior in the horizontal direction. However, for extreme heights, such as targets located directly above the user, the user's horizontal eye movement behavior may differ due to the user's inertia.

\begin{figure}[htbp]
\centering
\includegraphics[width=14cm]{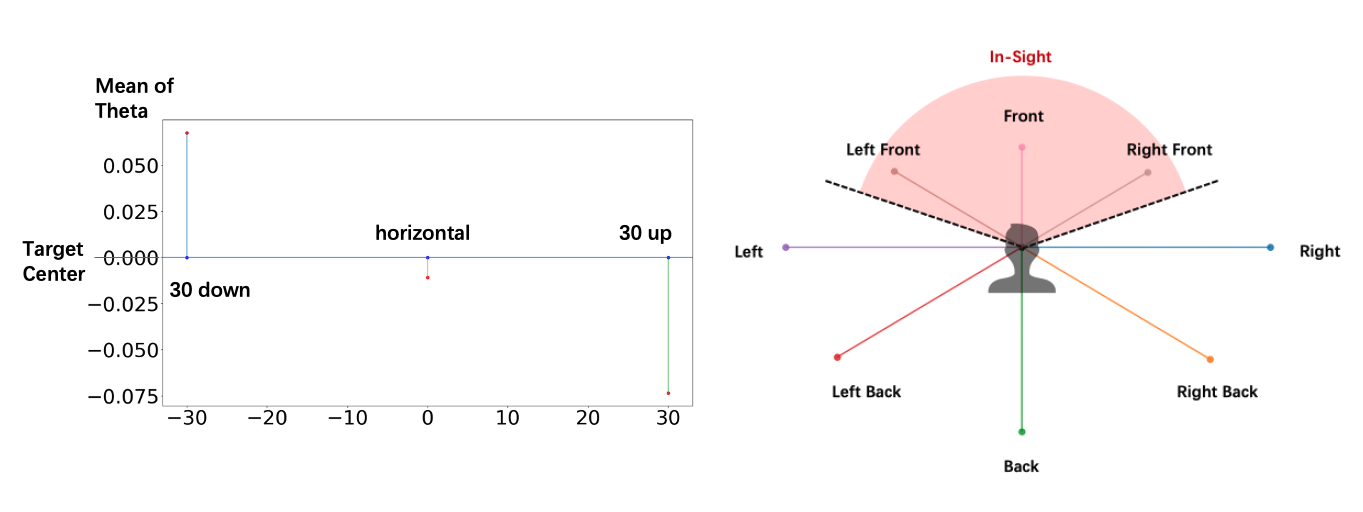}
\caption{Offsets at different heights (left). Black dots are target centers and red dots are offsets; Standard deviation of the distribution in different horizontal directions (right). The length of the line segment in different directions represents the value of the standard deviation.}
\label{fig_3_h_1}
\end{figure}

The horizontal angle affects the mean and variance in both the vertical and horizontal directions($\Theta{mean}:F_{(7,77)} = 5.309, ~p <0.05, \Theta{std}:F_{(7,77)} = 10.943, ~p <0.05, \Phi{mean}:F_{(7,77)} = 5.309, ~p <0.05, \Phi{std}:F_{(7,77)} = 18.231, ~p <0.05$). We found that users' eye movements also exhibit inertia, as the center of the gaze distribution on the left targets shifts to the right due to the user's gaze resetting to the front before the next trial. For the targets in the user's sight(front, front left, and front right), the user can easily adjust the gaze. When the target appears out of the user's sight, the user only has a vague impression of the target before making a selection. The standard deviation of the offset gets significantly increased. For targets located at extreme positions (behind, left behind, right behind), the user's vertical eye movement will also be affected, resulting in a significant deviation. Some users said, "I feel tired when I turn my body when looking at the target behind me, so I do not want to look at the target very accurately."



To simplify the construction of the behavior model, we focused on the selection data of the target in front of the user, as there are typically few interactions between the user and the targets behind them in real-life situations. The center of the Gaussian distribution is almost the same as the object's center. 
The radius of the distribution (17.18 degrees for in-sight targets) is much more extensive than the capture accuracy of the eye tracker, indicating that the error of the eye tracker does not cause the angular characteristics of the distribution.



\subsection{Part 2: Interaction Effect between two Nearby Objects for Casual Gaze}
In this part, we aim to explore how the presence of nearby devices can impact a user's gaze behavior. Our hypothesis is that the attributes of neighboring devices can influence the user's gaze behavior, even if they are not the intended target. 
A simple observation is that when two devices are placed near each other, the user may look far from the left device to indicate the right device as the target. 
To test this hypothesis, we conducted a controlled study in which participants were asked to select a target device from a pair of nearby devices.



\subsubsection{Participants and Apparatus}
We recruited 15 participants (eight males and seven females) with an average range of 23.4 (SD = 1.24) from the local campus via an online questionnaire. Twelve participants had prior experience with head-mounted VR displays. All of them had normal vision and did not participate in the first part of this study.

An empty room of $6 m \times 6 m \times 6 m$ was rendered in VR as the environment, with a pair of spheres with different parameter settings appearing in front of the room as targets. Other settings are the same as in the Part 1.


\subsubsection{Design}
We identified three independent factors: the distance between two nearby targets, the relative location of the targets, and the relative size of the target, which can affect the user's gaze point distribution. 

In Part 1, we found that the user's gaze distribution is not affected by the target size for an independent target. The independent target can be regarded as having an interference target far enough away from it. When the interference target is close enough, a shift in gaze may occur. We believe that the distance between targets can affect the gaze distribution. According to Part 1, the space radius occupied by the isolated target is 17.18 degrees. When the distance between the target and the user is 3 meters, the space radius is 0.92m. So we set four levels, 2 m (the distance cannot affect), 1.5m, 1m, and 0.5m (the smallest targets are close to each other), for the distance between the target centers. 

When neighboring targets' relative locations differ, the user's gaze distribution will also change. For instance, the distribution will shift in different directions for two nearby targets on the left and right. To account for this, we set eight relative locations for the target. For a specific target at the center, the surrounding targets will appear in eight directions, and the distances of these targets to the user will be the same (3 m). We only studied the situation of the relative location where there is no mutual occlusion. In occlusion cases, the user's gaze may have other changes besides the shift. We did not consider this in our study for simplicity.

In this part, the targets will be presented in pairs, consisting of a goal target and an interference target. As observed in Part 1, the goal and interference targets will have four different sizes, corresponding to the common objects in daily life. This results in a total of 16 possible combinations of target sizes.
Due to a large number of variables, we adopted an in-subject design to collect experimental data and ensure that each variable's overall collected data was balanced.


\begin{figure}[htbp]
\centering
\begin{minipage}[t]{0.3\textwidth}
\centering
\includegraphics[width=4.5cm]{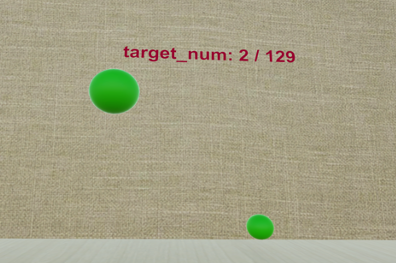}
\label{fig_22}
\end{minipage}
\begin{minipage}[t]{0.3\textwidth}
\centering
\includegraphics[width=4.5cm]{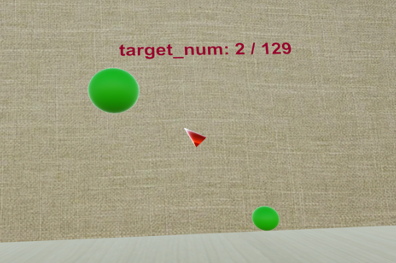}
\label{fig_23}
\end{minipage}
\begin{minipage}[t]{0.3\textwidth}
\centering
\includegraphics[width=4.5cm]{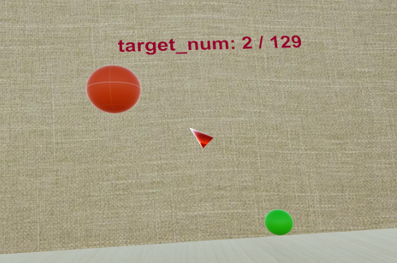}
\label{fig_23}
\end{minipage}
\caption{Task progress of Study 2. After the three-second countdown, two spheres appear in front of the user (left). The relative size, location, and distance of the two spheres are randomly generated. After confirming the location, the user pulls the trigger, a red cone appears and indicates to select the target (middle). The user select the target with gaze and pulls the trigger again. The selected target lights up to give the user feedback (right).}
\end{figure}
\subsubsection{Task and Procedure}
{An experimenter first explained the experiment to the participants.
The participants wore a headset and calibrated the gaze tracker with the experimenter's help. Before commencing the experiment, we conducted a demonstration program to familiarize users with casual gaze behavior patterns. The program featured two table lamps placed on a desktop. When users turn their heads or eyes to suggest one of the table lamps, the table lamp will light up. This process lasted for approximately five minutes until users reported that they could distinguish between the two lamps. This step was crucial in helping users understand the importance of casual interaction with the device rather than staring at it.

We informed participants to use casual gaze to suggest the target sphere. Each participant conducted 512 trials(4 sessions with 4 relative distances \* 4 actual target sizes \* relative target sizes \* 8 relative target locations) on target interaction. At the beginning of each trial, the participants were asked to reset their heads and eyes to the starting direction.
After a three-second countdown, two nearby devices with particular parameters appeared in the room, and participants had time to familiarize themselves with the targets. Upon triggering the handle, a small cone appeared in the center of the two targets, pointing toward the goal target. It was ensured that the participants did not solely rely on the direction of the cone but used casual gaze after learning the target information. As in Part 1, participants were required to press the trigger as soon as they felt that they had expressed their intention. Afterward, the participants reset their bodies and eyes to the starting direction. Each session's relative location, size, and distance are random for each participant. The experiment takes about 40 minutes, with a break every two sessions.
Each participant received a 16 USD gift card for compensation.

}

\subsubsection{Results}
We obtained 7740 selection data in the experiment, and we removed user-reported errors and large deviations in the overall statistics (109 times, 1.41\%).
We recorded the \emph{offset} between the user's casual gaze direction and the target position. However, it should be noted that some target combinations were not reachable at the lowest distance level, as the minimum distance between the two largest targets was 1m. Therefore, we will analyze the data in two parts. Firstly, we will consider the small size for all distances. Secondly, we will focus on the significant distance for all relative sizes.
We ran RM-ANOVA tests ($p < 0.05$) on the metric of offsets $(\Phi{s}, \Theta{s})$ with the variables of relative location, size, and distance, with the post-hoc T-tests ($p < 0.05$).
We highlight our major results and findings below. 

\textbf{1. Relative location causes a shift in the center of the distribution.} 
RM-ANOVA tests showed that the relative location has a significant effect on the offset in both directions($\Theta{mean}:F_{(7,98)} = 11.464, ~p <0.05, \Phi{mean}:F_{(7,98)} = 18.852, ~p <0.05$), and the shift direction is opposite to the direction in which the interfering target appears. 
When an interfering target appears in a particular direction, users tend to shift their gaze away from the interfering target to distinguish it from the goal target.
To further investigate the influence of the interfering target's direction, we decomposed the unit direction vector of the interfering target in the vertical and horizontal directions and studied their impact on the shift of the distribution center. We found that the angular offset in different directions was much larger than the error of the eye tracker, indicating that the offset was not due to error. Moreover, we observed that the shifts and component vectors were nearly linearly related in both directions.

When the distance between the interfering target and the goal target changes, the offset of the center of the distribution point does not change significantly($\Theta{mean}:F_{(3,42)} = 1.950, ~p =0.136, \Phi{mean}:F_{(3,42)} = 1.297, ~p =0.288$), and when the sizes of the targets changed, the center of the point distribution does not shift( $Goal target size: \Theta{mean}:F_{(3,42)} = 0.924, ~p = 0.437, \Phi{mean}:F_{(3,42)} = 0.816, ~p =0.492$). These findings suggest that the user believes they can express the intention to distinguish selections as long as they shift their gaze to the target direction. When the interfering target's size and distance change, the offset's size will not be significantly affected.
\begin{figure}[htbp]
\centering
\includegraphics[width=\textwidth]{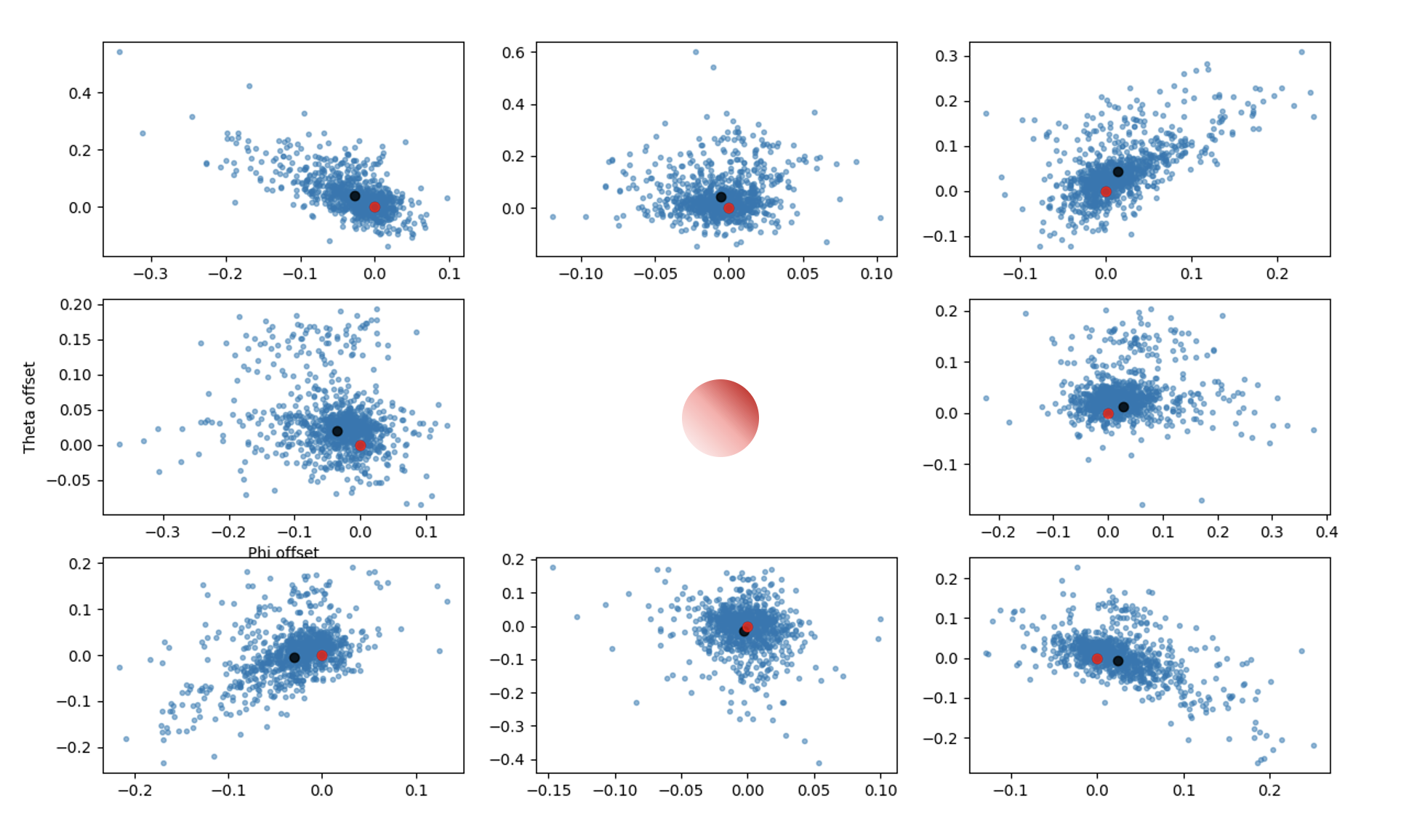}
\label{fig_23}
\caption{The distribution under different relative locations. Different colors represent the Gaussian distribution under different relative locations. The black point represents the center of the Gaussian distribution after the shift. It shows that the relative location causes a shift in the center of the Gaussian distribution, and the shift levels in the two directions are different.}
\end{figure}


\textbf{2. Relative distance and size influence the radius of the distribution.} 
RM-ANOVA tests showed that the relative distance has a significant effect on the standard deviation in both directions($\Theta{std}:F_{(3,42)} = 17.096, ~p <0.05, \Phi{std}:F_{(3,42)} = 7.311, ~p <0.05$), and the relative location also has a significant effect on the standard deviation of the distribution($\Theta{std}:F_{(7,98)} = 5.857, ~p <0.05, \Phi{std}:F_{(7,98)} = 10.404, ~p <0.05$). The change in the target's location means the change in the distance in both the horizontal and vertical directions. When the interference target keeps getting closer, 
the user will more accurately look at the target to express selection intention clearly, which means the user's gaze will be close to the selection target. Therefore, the standard deviation of the distribution will gradually decrease. The closer the distance to the interference target, the more concentrated the user's attention will be. We fit the size of the standard deviation and distance and find that the two factors are linear related.

The size of the target itself has a significant impact($\Theta{std}:F_{(3,42)} = 3.960, ~p <0.05, \Phi{std}:F_{(3,42)} = 13.255, ~p <0.05$) on the standard deviation. In contrast, the size of the interference target has no significant impact($\Theta{std}:F_{(3,42)} = 0.542, ~p =0.656, \Phi{std}:F_{(3,42)} = 0.925, ~p=0.437$). Interfering targets can compress the user's gaze distribution toward the selection target. The degree of compression is influenced by the size of the target, with smaller targets resulting in a smaller distribution radius. This is because the user's gaze tends to concentrate near the target. 
However, changes in the size of surrounding targets do not significantly affect the distribution, as it remains concentrated near the selection target. Additionally, we conducted a regression analysis to examine the relationship between the standard deviation and target size. Our results show a linear relationship between the two variables.

The distribution of gaze points on a target can be influenced by the presence of other targets in the vicinity. Notably, these effects tend to be linear in nature. When information about multiple targets is available, it is possible to estimate the distribution by analyzing the fitting results.
\begin{figure}[htbp]
\centering
\includegraphics[width=\textwidth]{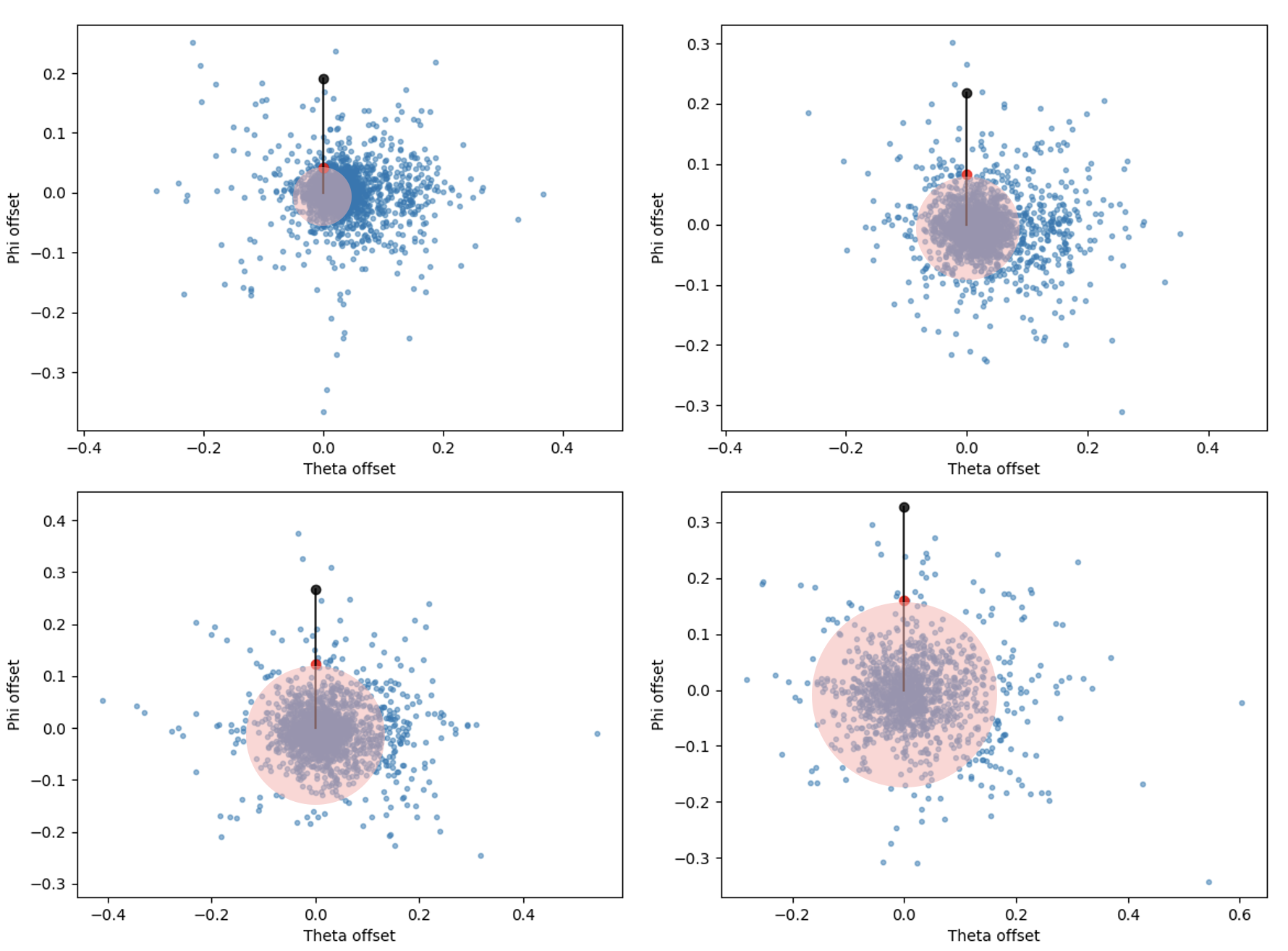}
\label{fig_23}
\caption{Gaussian distribution of different goal target sizes. The red circles represent the target itself, the red dots represent the target edges, and the black dots represent the radius of the Gaussian distribution.}
\end{figure}
\section{Study 2: Investigating Casual Gaze Behavior in Real-World Environments}
After obtaining the effects of different target parameters through the study of independent objects and pairs of objects, we further conducted a study of the user's casual gaze behavior in a simulated real-world environment. We used VR to simulate the environment, which had a similar layout to a real-world room with distributed smart devices, to explore the user's casual eye movement behavior. This study aimed to 1) provide analysis of the user's eye movements to solve the ambiguous region problem that appeared in the previous experiments and 2) collect casual gaze data in a simulated real-world environment to facilitate further simulations on the recognition algorithms.

\label{sec:stu_3}
This study studied the interaction between multiple devices in a smart environment. There are often multiple devices with different locations and characteristics in the real space. 
To achieve this, we built a room with 18 devices in virtual reality and collected data on users' casual gaze to express their selections. 
We constructed a series of Gaussian models using the user's gaze to describe the distribution in a specific environment. These models are a particular case of the environment we build. We recorded the user's gaze behavior throughout the process to facilitate the analysis of the temporal characteristics of casual gaze.


\subsection{Participants}
{
We recruited 16 participants (eight males, eight females) with an average range of 23.25 (SD = 1.39) in this experiment. 13 participants had prior experience with head-mounted VR displays. They all had normal vision and did not participate in the prior study.
}

\subsection{Apparatus}
{ 
A smart room of $6 m \times 9 m \times 3 m$ was rendered as the environment, with 18 devices as the target appearing at different locations in the room. Other settings are the same as in Study 1.
}

\subsection{Design}
{In this study, we set up a room with different smart devices in virtual reality. To simulate the real scene as much as possible, we chose 18 different daily equipment (speaker $\times$ 3, refrigerator $\times$ 1, air conditioner $\times$ 1, sweeping robot $\times$ 2, laptop $\ times$ 2, electronic clock $\times$ 1, electric light $\times$ 6, washing machine $\times$ 1, TV $\times$ 1). Each device is marked with a name and a number for users to remember and distinguish. Our selection criteria were based on the devices commonly used in daily life and frequently interacted with by users. To guarantee the richness and versatility of the data, we also ensured that the devices covered a wide range of special cases.

We divide the current devices into five cases according to its own parameters and overall layout. These situations present different challenges for gaze selection: 
\begin{itemize}
\item Normal(N): the device is neither too large nor too small, and is located at a moderate distance from other devices. The device is located in front of the user, making it easy to select using gaze. 
\item Small view occupied(S): The device size is small or the long distance from the user makes the device small in the user's sigh.
\item Special located(L): The device is located where it is inconvenient to interact using the gaze (e.g. behind or above the user).
\item Close proximity(C). There is one or more other devices of similar size and proximity to the target.
\item Disparity in size(D). There is one or more other devices with a significant size difference near the target.
\end{itemize}

The location of devices is the same as in daily use and covers all five cases above. The clock is on the wall in front of the user. The TV and washing machine are placed in the room in front of the user. Two speakers are placed on the left and right sides of the TV, and another speaker is placed on the bookshelf behind the user. The refrigerator is in the left corner facing the user, and the air conditioner is on the wall behind the user. Two sweeping robots are placed on the ground. Two laptops are placed on the user's left and right desks. Six electric lights are installed on the ceiling. These devices cover the possible locations of daily devices.
The device size is the same as in the actual scene and covers all five cases above. These sizes ranged from small electronic clocks and speakers to larger appliances such as refrigerators and washing machines, covering all possibilities of common devices.

\begin{figure}[htbp]
\centering
\begin{minipage}[t]{0.48\textwidth}
\centering
\includegraphics[width=7cm]{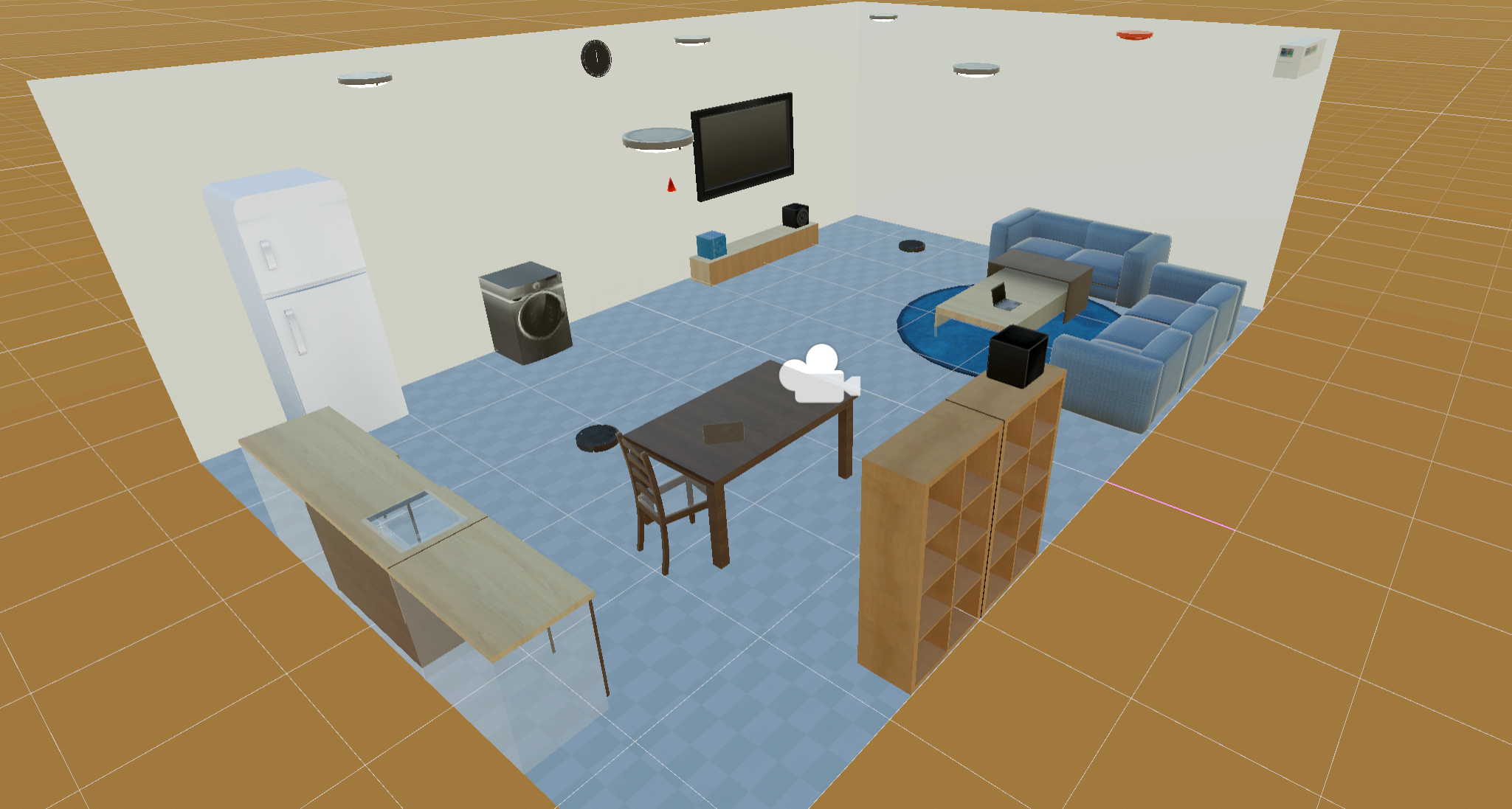}
\label{fig_22}
\end{minipage}
\begin{minipage}[t]{0.48\textwidth}
\centering
\includegraphics[width=7cm]{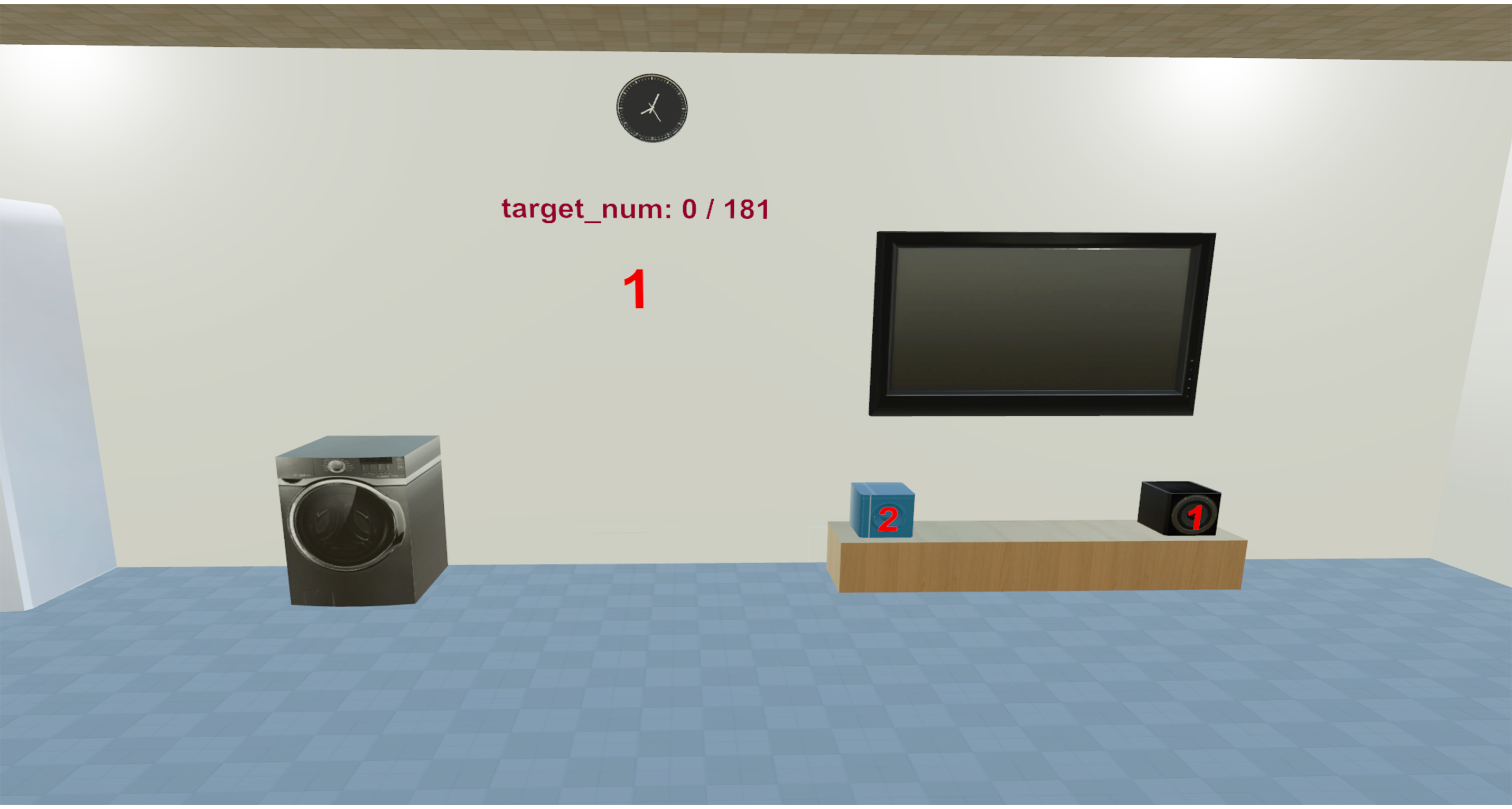}

\label{fig_23}
\end{minipage}
\caption{Environment settings for Study 3. There are 18 different smart devices in the environment. We placed them in different locations according to the actual situation and ensured that the relative locations and relative distances between them were diverse. The figure on the right is the user's first perspective.}
\end{figure}
}

\subsection{Task And Procedure}
{An experimenter first explained the experiment to the participants.
The participants wore a headset and calibrated the gaze tracker with the experimenter's help. Before the experiment, the experimenter will guide users to familiarize themselves with the entire experimental environment, including the device's location, name, and number. The participants had time to remember the device information. This process continues until the user reports that they are familiar with the environment, which takes about 10 minutes.

Each participant conducted 720 recommended equipment trials (4 $\times$ 180 trials). At the beginning of each trial, the participants reset their bodies and eyes to the starting position. After two-second countdown, the name and number of the target will appear in the user's sight. The users were instructed to use a casual gaze to interact with the device and trigger a button on the controller when they felt that their intention had been expressed. Then, the user reset the head and eyes to the starting direction. We require users to express their intentions naturally, accurately, and quickly.

The order in which the target devices appeared was randomized in each session. The entire experiment took approximately 60 minutes, with a break provided after every two sessions.
Each participant received a 16 USD gift card as compensation.

}

\subsection{Results}
A total of 2781 valid data were collected in the experiment. We recorded the user's gaze direction and target information during the experiment. We calculated the \emph{offset} of the user's final gaze and target position. 
We highlight our significant results and findings below. We use the collected data to fit the two-dimensional Gaussian model for 18 devices. These models are established for the user's casual gaze behavior in this specific scenario. The model will be used for the validation of subsequent behavioral model inference. 
We highlight the findings on eye movement behavior below.

%


\textbf{1. Speed and randomness of gaze increase with familiarity.}
Each user performed four sessions of the selection task in the experiment. The user's familiarity with the environment increased as the task progressed. We analyzed the speed and device distribution of users' selection in each session. We found that as familiarity increased, the user's selection speed gradually accelerated, and the distribution radius also increased. This suggests that the user's gaze was more casual in the familiar environment. Although the experimenter provided guidance to help users become familiar with the environment, the user still needed time to memorize the device layout and information. At the beginning of the task, users had concerns about selecting the wrong target and tended to look more accurate. However, users became more familiar with the devices' location and size as the experiment progressed. They tended to use casual gaze to indicate the target with a lower burden. 
\begin{figure}
\centering
\includegraphics[width=10cm]{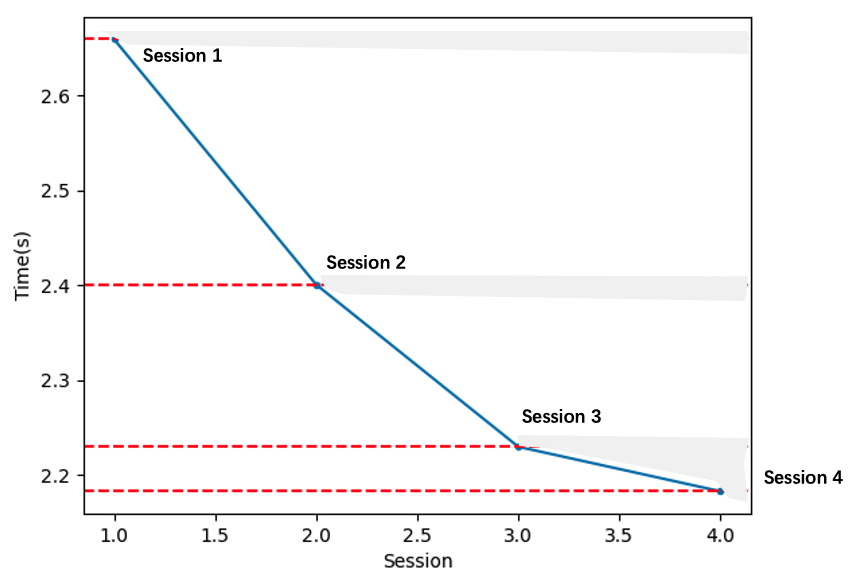}

\caption{Average user selection time for different sessions. As familiarity increases, the user's selection speed keeps increasing.}
\end{figure}

\textbf{2. Eye movement has special trends in blurred areas.}
Casual gaze has ambiguous regions between devices. It is difficult to identify the user's intention using only the final eye movement direction, so we visualize the gaze trajectories falling in ambiguous regions to understand the casual gaze behaviors. Since casual gaze does not require the user's gaze to fall within the target, some gaze behaviors are inevitably ambiguous. 

\begin{figure}
\centering
\includegraphics[width=0.8\textwidth]{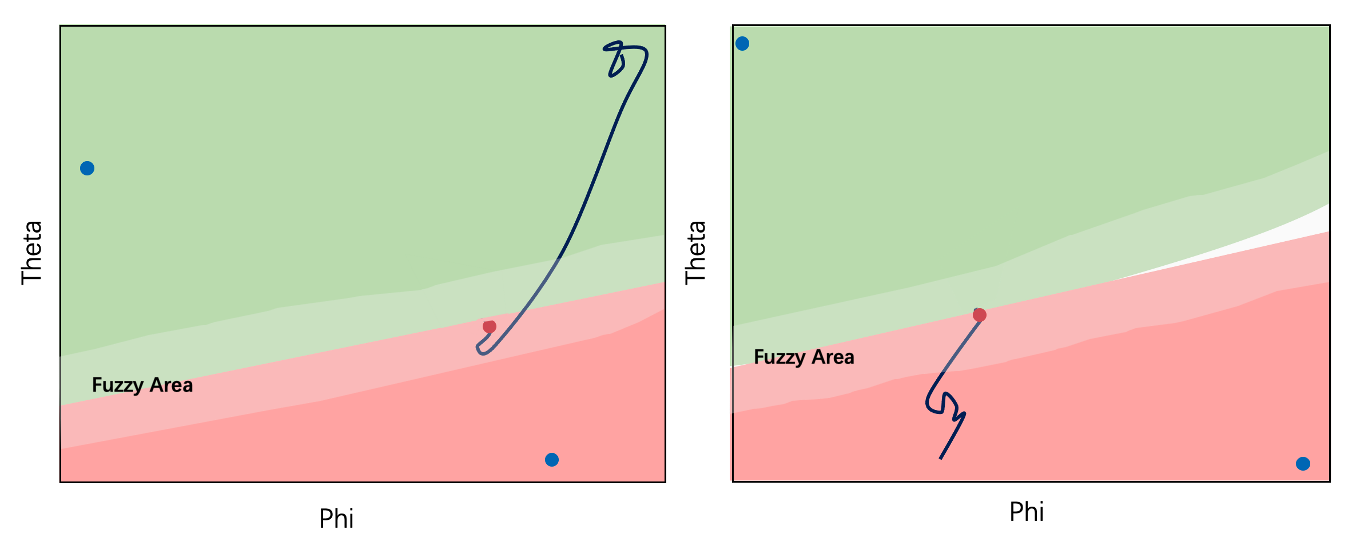}

\caption{The gaze of the fuzzy region. Red and green areas represent the Gaussian distribution of the adjacent devices. The blue line is the user's gaze trajectory, and the red point is the final gaze.}
\end{figure}

However, two types of behaviors can still be resolved by trajectory features. The first type of eye movement is too fast, causing the gaze crosses the target and falls into the blurred area. Typically, there is usually a foldback at the end of the trajectory. The second type is the early deceleration of the eye movement, causing the gaze to fall into the blurred area before reaching the target. In this case, the gaze trajectory will generally pass through other targets. These two trajectories provide valuable information about the temporal order of selection intention. The closer to the end of the trajectory, the greater the value of the gaze. Partially ambiguous behaviors can be judged by predicting the user's behavior with the end segment of the trajectory. 
However, users may need to confirm or adjust their gaze for inherently ambiguous behaviors to complete the target interaction. By visualizing eye movement trajectories, we can better understand casual gaze and improve the accuracy of interpreting user intent.

\section{CasualGaze Algorithms}
\label{sec:model}



\begin{figure}[htbp]
  \includegraphics[width=0.9\textwidth]{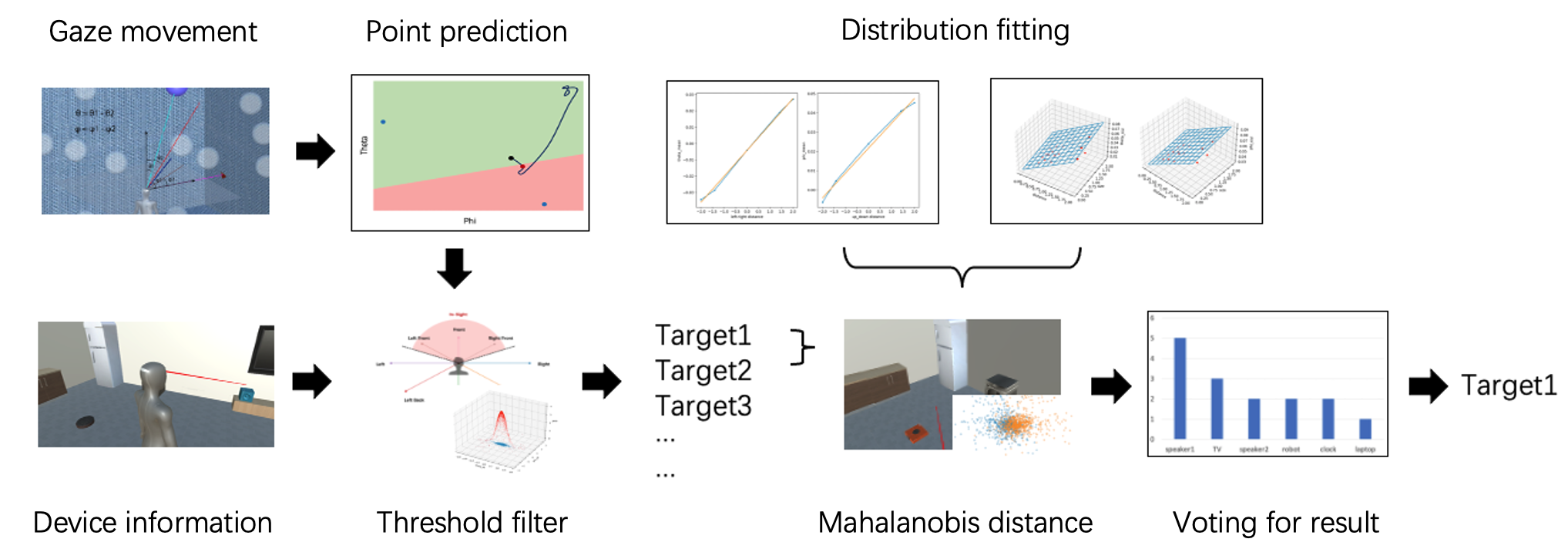}
  \caption{The formula of the Gaze model. The model first predicts the gaze point based on the user's eye movement, and then gets candidate devices based on the data of Study 1-1, compares every two candidate devices according to the fitting result of the Gaussian distribution, and finally votes to obtain the prediction result of the target device.}
  \label{fig:set}
\end{figure}
Based on previous studies, we obtained the casual gaze distribution around target and trajectory behavior. 
We established a behavior prediction model according to the study results and performed curve-fitting on critical variables in the model to ensure they could be applied to more common situations. The predictive model provides a Gaussian model (Equation \ref{equ_1}, According to the study results, the covariance is 0) of each device in the user's sight based on parameters that significantly impact user behavior. The model predicts the user's gaze and uses the offset between the gaze and the device position as an input to implement intent calculations. Once the device's information is known, our model can predict whether the user is interested in the device in the scene based on the user's gaze and further determine the device with the highest user interest.
\begin{center}
\begin{equation}
\label{equ_1}
\begin{aligned}
f(X)&=\frac{1}{2\pi|\Sigma|^{\frac{1}{2}}}exp[-\frac{1}{2}(X-u)^{T}\Sigma^{-1}(X-u)],X=(\theta,\phi)\\
u &= (u_{\theta}, u_{\phi}), 
\Sigma  = 
\begin{bmatrix}
 \delta_{\theta} & 0.0\\
 0.0 & \delta_{\phi}
\end{bmatrix}
\end{aligned}
\end{equation}
\end{center}

We first give a list of devices that users may be interested in based on Study 1-1. Then we build a dual-device distribution compression prediction model based on Study 1-2. We predict the user's gaze from the trajectory and compare every two devices with a higher probability of getting attention based on the gaze. We finally use a voting algorithm to calculate the final goal.

\subsection{Speed Weighting and Gaze Prediction}
According to the trajectory of the user's gaze, we found that the user's gaze movement describes the user's trend of selecting the target. We analyzed the user's gaze movement in the last 0.2s and tried to predict the actual gaze direction. We chose movement vectors based on the five frames of data in the last 0.2s. The vector between every two adjacent frames represents the direction and speed of the user's eye movement. Considering that the closer to the current gaze, the higher the value of the data, we selected the weights of increasing arithmetic and weighted these vectors to obtain the prediction of the direction of gaze. Let ($t_{p}, p_{p}$) be the predicted gaze, they can be calculated with the data of the last 5 frames.
\begin{center}
\begin{equation}
(t_{p},p_{p}) = (t_0,p_0) + \sum \left [(t_i,p_i) - (t_{i-1},p_{i-1})\right ]\times k_i, i = 0, -1, -2, -3, -4.
 \end{equation}
\end{center}
where $t_0$, $p_0$ indicates the user's current gaze, k is the weighting coefficient associated with i.

\begin{figure}[htbp]
  \includegraphics[width=0.8\textwidth]{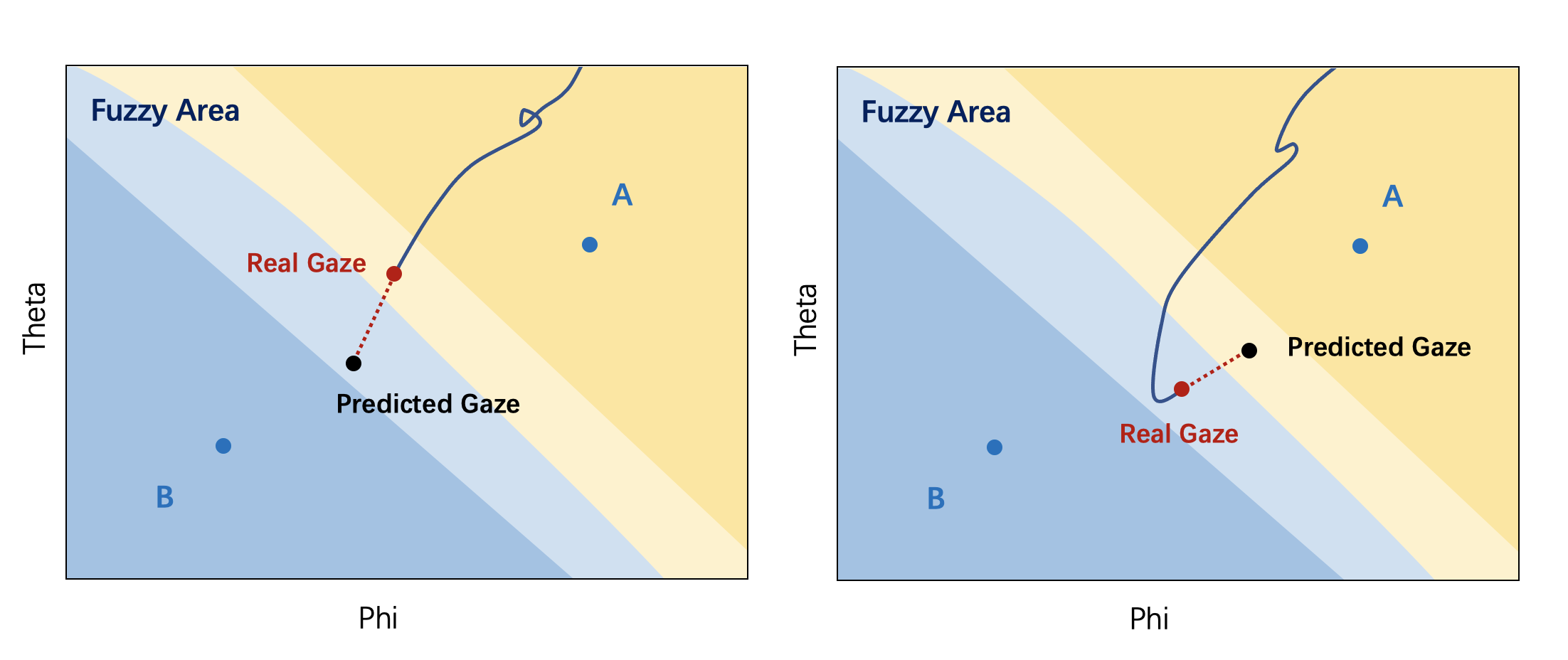}
\caption{The case when gaze point falls on the boundary. The blue line represents the trajectory of the gaze, and the red dot above it represents the actual point of sight of the user. We weighted the results of several consecutive frames according to the speed and direction of the gaze movement and predicted the user's possible target point (black). }
\label{fig:set}
\end{figure}

\subsection{Interested device set generation}
A device with the head facing less than 96 degrees from the forward head direction($F_h$) is in the user's field of view. This is because the user's standard horizontal vision is 130 degrees, and their vertical vision is 120 degrees (50 degrees upward, 70 degrees downward). 

Based on Study 1, we can determine whether a user has potential interaction intentions for a device. Specifically, we found that when selecting a separate interactive object in front of them, the offset of the user's gaze direction($G$) from the target does not exceed 17.18 degrees. It's worth noting that we did not consider the shape and appearance of the device in this determination. 

We add devices that satisfy both conditions to the candidate set. 
Let $P_h$ denote the position of the user's head, $P_e$ denote the position of the user's eyes, and $P_t$ denote the target position. The conditions to be satisfied are given below.

\begin{center}
\begin{equation}
\begin{aligned}
Angle_{1} &= 
\arccos \frac{F_{h}\cdot (P_{t} - P_{h})} {\left | F_{h} \right | \left | P_{t} - P_{h} \right | }  < 96^{\circ} \\
Angle_{2} &= \arccos \frac{G\cdot (P_{t} - P_{e})}{\left | G \right | \left | P_{t} - P_{e} \right | } < 17.18^{\circ} 
\end{aligned}
 \end{equation}
\end{center}

After obtaining the set of candidate devices, we simplify the interference between multiple devices. To achieve this, we will consider the distributional influence of other devices on specific devices separately. We fit distributions on each pair of Candidate devices for further intent recognition.



\subsection{Distribution fitting of paired devices}

Based on the set of candidate devices obtained, we will analyze each pair according to the conclusions of Study 1-2.
Parameters that have an impact on the distribution include device size, relative location, and distance. We fitted the mean and standard deviation of the distribution based on these parameters. Before performing the fitting, we normalized all parameters with a reference distance of 3m from the user.

The two devices' relative position affects the distribution center's horizontal and vertical offset. The shift and angles are nearly linearly related, and the degree of influence in the vertical and horizontal directions is different. We performed the least-squares linear fitting on the data in Study 1-2 in these two directions. We have recorded the parameters of two straight lines and can predict the position of the distribution center  when the relative position of the two devices is given. Using ($\phi_{t}, \theta_{t}$) to represent the angular coordinates of the target and ($\phi_{d}, \theta_{d}$) to represent the angular coordinates of the disturb target, the fitting formula is as follows.

\begin{equation}
\begin{aligned}
\phi_{mean} &= a_{\phi}\times (\phi_{t} - \phi_{d}) + b_{\phi} \\
\theta_{mean} &= a_{\theta}\times (\theta_{t} - \theta_{d}) + b_{\theta} 
\end{aligned}
\end{equation}

, where $a_*$ and $b_*$ are the fitting parameters.

\begin{figure}
  \includegraphics[width=0.8\textwidth]{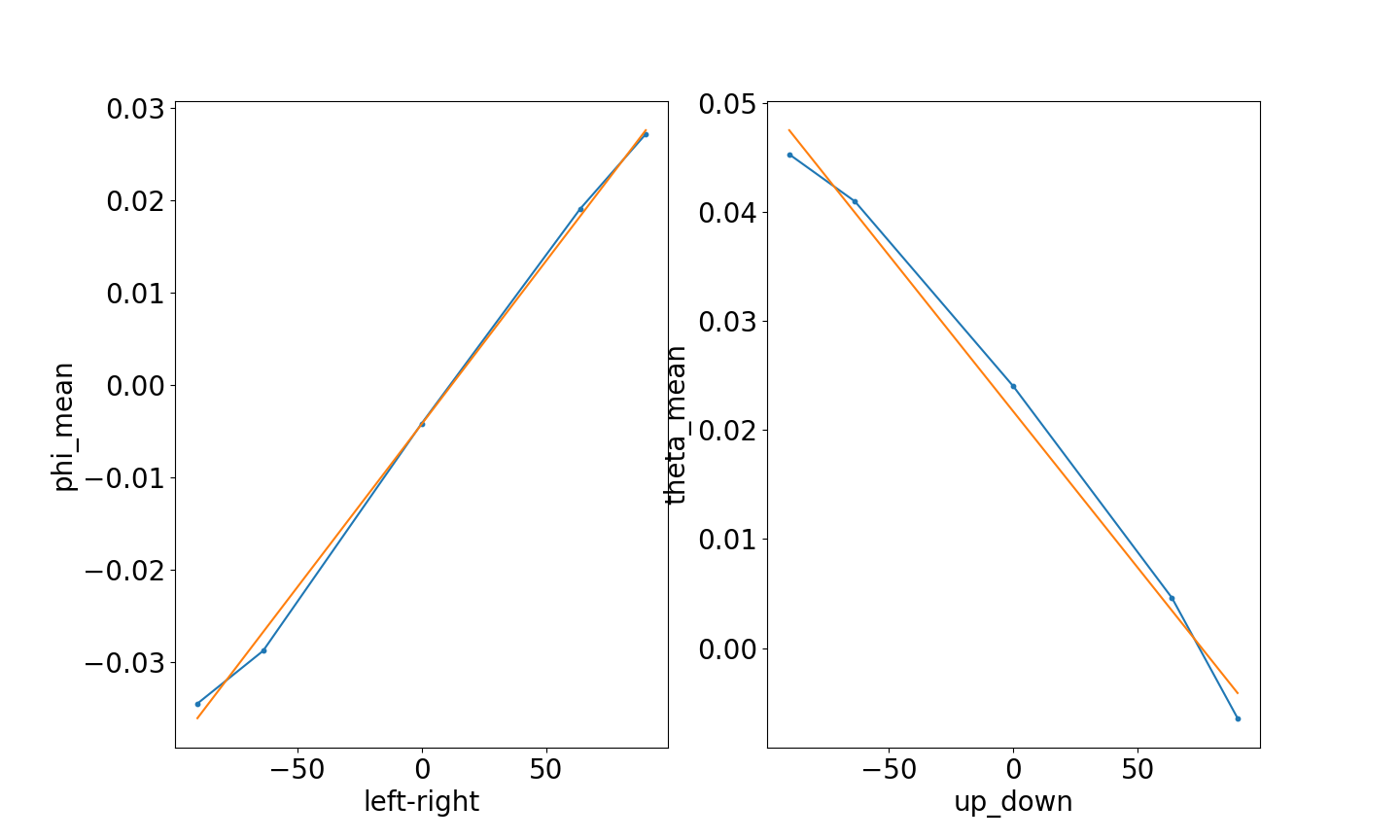}
  \caption{The influence of the relative position vector on the center of the distribution. The least-squares method is used to fit the vector's influence curve on the offset of the center in two directions. When a relative position is given, the corresponding offset of the center in the two directions can be calculated respectively.}
  \label{fig:set}
\end{figure}

For the standard deviation of the distribution in the two directions, we found that it is significantly related to the size of the target and the relative distance to other targets. The influence of these two factors on the standard deviation is linear, and their influence is independent. So we take these two variables as independent variables and the standard deviation as the dependent variable to fit a plane. Since the distance between the target and the user in Study 1-2 is 3m, the distance between the two devices and the target size should be standardized in the spherical coordinate system. Using $S_{t}$ to represent the origin size of the target, $P_{t}$ 
to represent the origin position of the target, $P_{e}$ to represent the origin position of the user's eye, ($\phi_{t}, \theta_{t}$) to represent the angular coordinates of the target and ($\phi_{d}, \theta_{d}$) to represent the angular coordinates of the disturb target, the fitting formula is as follows.

\begin{equation}
\begin{aligned}
\phi_{std} &= a_{\phi} \times  S_{t}\times\frac{\left | P_{t} - P_{e} \right | }{3} + b_{\phi} \times ( 3 \times \tan\frac{\left | \phi_{t} - \phi_{d} \right |}{2}\times 2)  + c_{\phi} \\
\theta_{std} &= a_{\theta} \times S_{t}\times\frac{\left | P_{t} - P_{e} \right | }{3} + b_{\theta} \times ( 3 \times \tan\frac{ \left | \theta_{t} - \theta_{d} \right |}{2}\times 2) + c_{\theta}
\end{aligned}
 \end{equation}

 , where $a_*$, $b_*$, and $c_*$ are the fitting parameters.

\begin{figure}
  \includegraphics[width=0.8\textwidth]{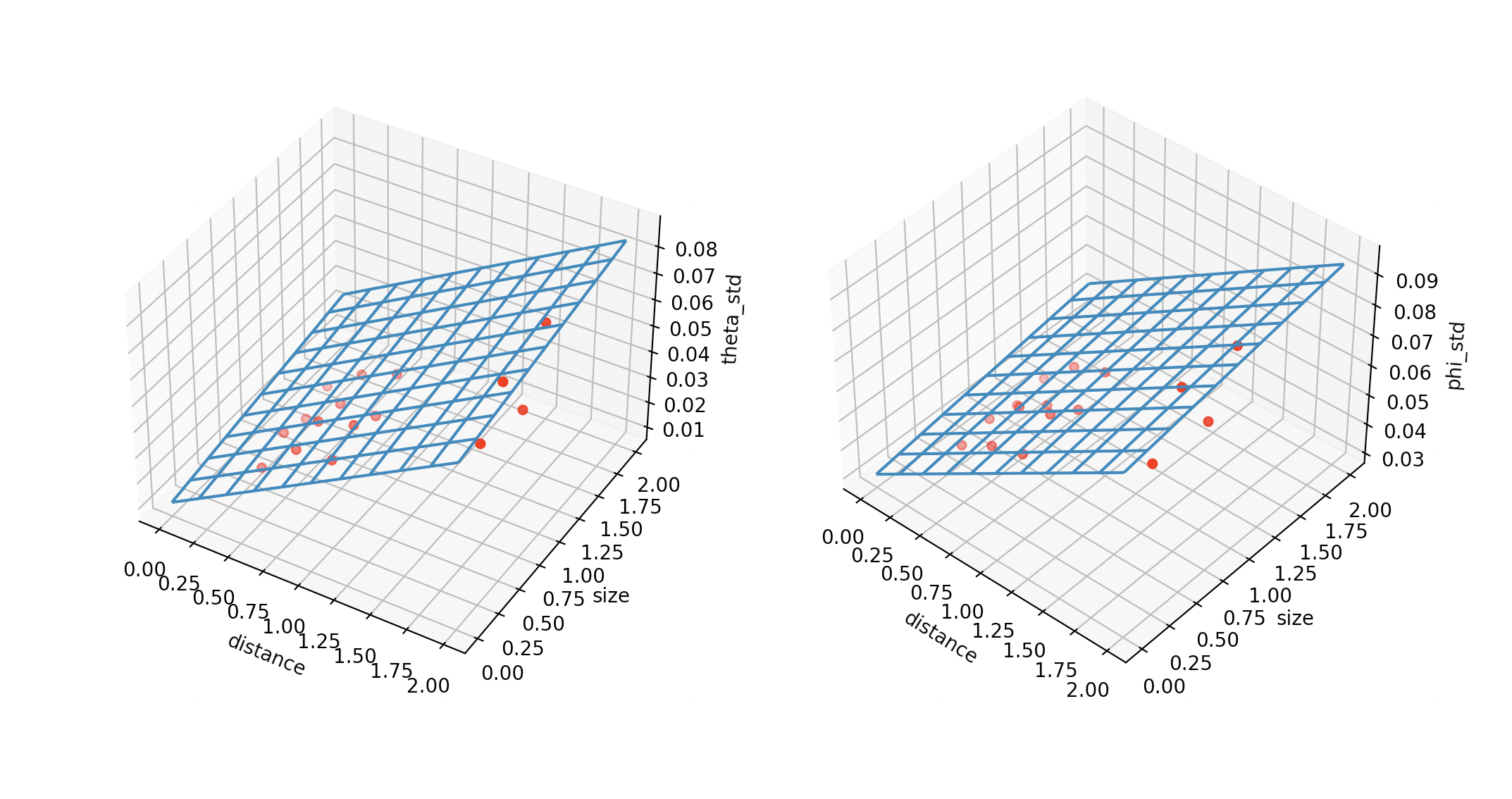}
  \caption{The influence of relative distance and target size on distribution radius. The relationship between the distribution radius and the two independent variables is close to linear. When a set of relative distances and target sizes are given, a curved surface fitted by the least square method can be used to calculate the corresponding radii distributed in two directions.}

  \label{fig:set}
\end{figure}

So far, when two devices information is given, we can obtain the Gaussian probability density distribution model of the casual gaze around the two devices according to the two straight lines and two planes obtained by the fitting.

\subsection{Voting Based Target Prediction Mechanism}
For each pair of devices, we can obtain the user's preference for these two devices. We adopt an algorithm based on the voting mode to select the device with the most significant user attention. We calculated the Mahalanobis distance between the predicted gaze and the distribution of the two devices\ref{dm} to compare the user's attention to the two devices. The specific operation is to initialize the target attention in all candidate sets to 0. When the device wins a pair of comparative analyses, its attention will get one vote. Finally, after completing each pair of comparative analyses in the to-be-selected set, the device with the highest number of votes is the device with the highest user interest calculated by the model.
\begin{center}
   \begin{equation}
   \label{dm}
       D_{M}(x) = \sqrt{(x - \mu )^{T}\Sigma ^{-1}  (x - \mu ) }  
    \end{equation} 
\end{center}

\begin{center}
   \begin{equation}
Score(1) += \left\{\begin{matrix}
 0,  D_M(1) >  D_M(2) \\
 1,  D_M(1) \le D_M(2)
\end{matrix}\right.
\end{equation} 
\end{center}

\subsection{Performance Evaluation}
In Study 2, we collected data on users' eye-movement selections in smart scenes and calculated the Gaussian models. Based on this data, we evaluated the \projectName{} performance. However, we also wanted to compare the performance of the K-Nearest Neighbor (KNN) approach, which is a more straightforward method that does not require eye movements to fall exactly within the target. We compared the specific Gaussian model, KNN, and \projectName{} to discuss their performances.

\textbf{1. \projectName{} has good generalization.} 

We calculated the accuracy of the three methods for user intent recognition. Our results showed that the specific model had an accuracy of 96.81\%, KNN had an accuracy of 95.94\%, and \projectName{} had an accuracy of 96.47\%. Even the specific model had a recognition error rate of 3.19\%, caused by ambiguous regions. Notably, \projectName{} demonstrated a recognition ability that was very close to the specific Gaussian model and higher than the KNN method. We simplified the situation of multiple devices by regarding it as the superposition of the effects of two devices. Although the interaction between multiple devices may be more complicated, the model based on current research already has good generalization. With a given room setting, \projectName{} can accurately predict the device that the user is interested in without pre-data training.

\textbf{2. \projectName{} is more adaptable than KNN in special cases.} 
\label{res:stu_3_3}
Regarding the algorithm, KNN is a specific case that does not take into account changes in distribution. This means that KNN assumes that the gaze distribution of all devices is consistent. On the other hand, \projectName{} predicts the distribution based on device parameters and layout, and incorporates the prediction of gaze behavior. This makes \projectName{} a more robust and accurate algorithm compared to KNN.



We discuss the special case presented in Study 2 and find that for the Disparity in size (D) case, \projectName{} has the most significant accuracy improvement with respect to KNN. The accuracy of \projectName{} is 97.61\%, while the accuracy of KNN is 95.29\%. We take the cleaning robot next to the refrigerator and the speaker next to the TV as examples.
The predicted boundary of KNN is still in the center of the distance between the two devices. However, according to Study 2, the gaze distribution of cleaning robot and speaker is more compressed, so the user's psychological boundary will be closer to small devices. When the user's gaze falls close to the boundary, the prediction results of KNN and \projectName{} will be different.

\projectName{} is more in line with the user's natural behavior and psychological model. Although \projectName{} does not have an exaggerated accuracy improvement, its performance in an interactive experience and complex scenes will improve much. We will analyze the performance of different techniques under different device cases in the Study 3.

\textbf{3. \projectName{} is more suitable for familiar scenes.} 
The analysis from Study 3 shows that as users become more familiar with the environment, their eye movements become more casual.
We analyzed the prediction results of the three methods in the four sessions. As the familiarity increases, the user's gaze has a more significant deviation from the target, so the recognition accuracy of the three methods decreases slightly.
It shows that the model of \projectName{} aligns more with the user's psychology. As users become more familiar with the environment, the accuracy of \projectName{} is getting closer to the specific Gaussian model, while the accuracy of KNN is still low. As the user's familiarity with the scene increases, their behavior is closer to the natural and casual eye movement represented by \projectName{}.
\begin{figure}
  \includegraphics[width=0.8\textwidth]{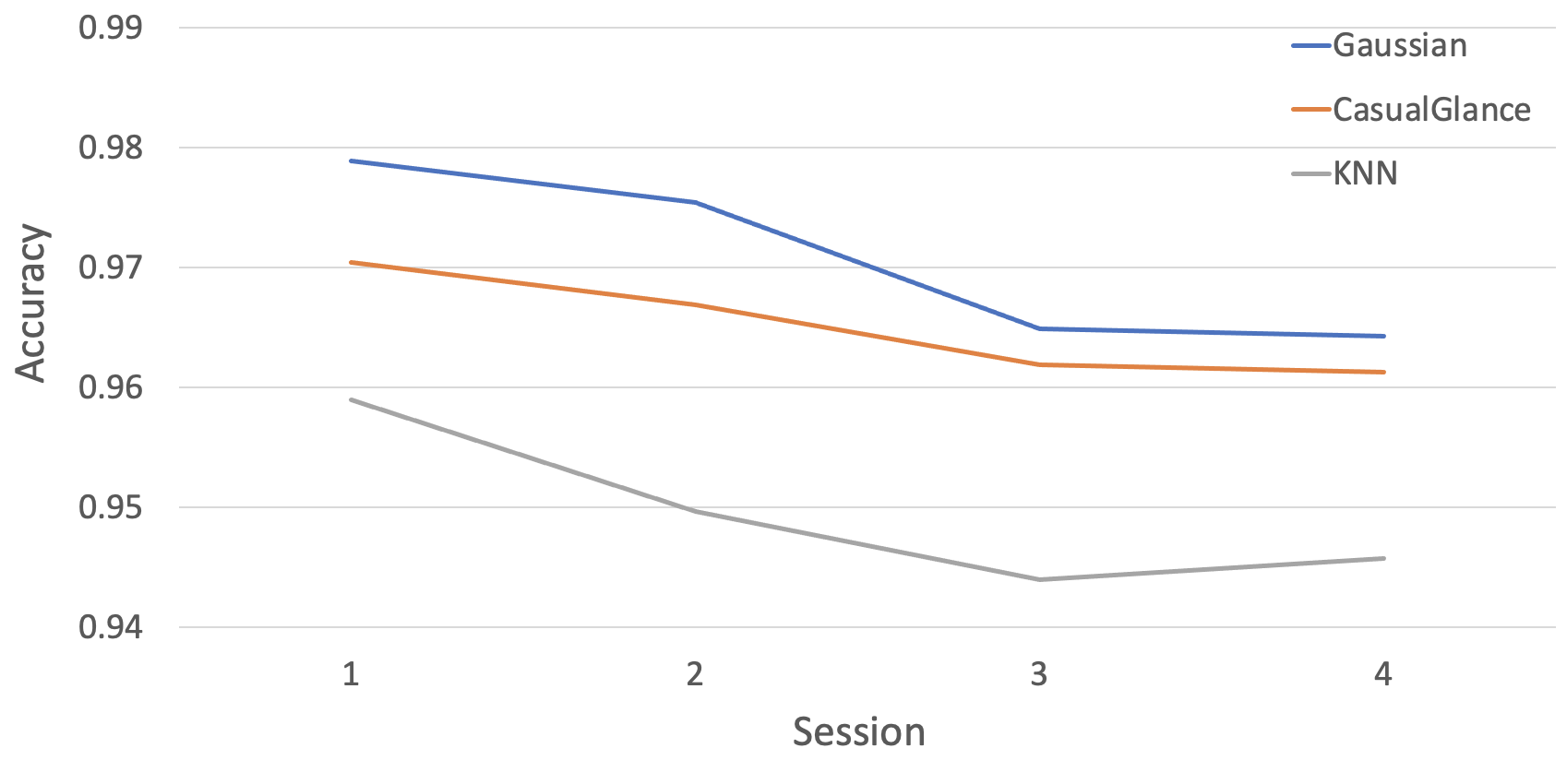}
  \caption{The accuracy of three technologies in different sessions. As users become more familiar with the environment, the overall offset of gaze relative to the target gradually increases, the accuracy of \projectName{} gradually approaches that of the Gaussian model, and the accuracy of KNN remains at a low level.}
  \label{fig:set}
\end{figure}

\section{Study3: Usability Evaluation of CasualGaze}
\label{sec:stu_4}
In this study, we tested the real-time performance of \projectName{} in user intention recognition. We selected three actual scenes in daily life and placed different types of devices in the scenes. In order to enhance the authenticity, we scanned the natural environment and migrated to the virtual reality environment. Users will use three methods to complete device selection tasks in different scenarios. In each scenario, we compared \projectName{}, KNN, and the baseline method that requires the user to stare at the device accurately. During the task, we collected data to compare the speed and delay. We interviewed users at the end of the experiment to obtain the users' feelings about using the three methods,  finally getting their subjective evaluation. We will compare the improvement of \projectName{} to the baseline method from objective data and subjective experience.

\subsection{Participants}
{
We recruited 13 participants (8 males, 5 females) with an average range of 22.46 (SD = 1.78) in this experiment. Nine participants had prior experience with head-mounted VR displays.  They all had normal vision and did not participate in the prior study.}

\subsection{Apparatus}
{
With the Lidar of iPhone 12 pro, we use Polycam to scan three actual scenes and import them into virtual reality as experimental scenes. Other settings are the same as in Study 1.}

\subsection{Design}
We evaluated three methods: baseline, KNN, and \projectName{}. The baseline method requires the user's gaze to fall accurately on the device, the same as looking at the device in common gaze interactions. This method does not consider the situation where the user's gaze has offset with the devices but still represents the user's selecting intention. The KNN method calculates the distance between the user's gaze and the device in the two-dimensional coordinate system of the angle component in real time and regards the device with the shortest distance as the user's target. KNN is the most direct method when the user's gaze has offset with the device, and it has achieved good results in Study 3. \projectName{} is the casual gaze model from user natural behavior.

We provided participants with three common scenarios (living space, simple working space and complex working space) in the study. In daily life, users are more casual in the living environment. In the work scene, there are more densely located devices that can cover possible device distribution. At the same time, two levels of target number are set in the work scene to analyze the performance under different intensity. In order to ensure naturalness and authenticity, each scene was scanned from the real scene with Lidar. 

There are 12 devices in the living scene, 
20 devices in the office scene. 
In the working scene with less equipment, we removed some computers and lights, reducing the number of equipment to 10.
The location and size of the equipment in all scenes are consistent with the actual situation of the scene. After scanning, it is imported into the virtual reality environment in its original size.

We named each device related to the location to help users quickly remember the devices and their locations. We also invited some users familiar with the working environment to participate in the experiment to obtain casual gaze behavior data.

We rendered a bounding box for each device in the virtual reality environment to provide feedback. 
In each scene, users use three different methods to interact with the devices. The method will continuously calculate the target device during the experiment and highlight the box to provide visual feedback. The sequence of scenes and methods are randomly generated among different users.

We have no requirements for the user's posture, and the user can complete the experiment sitting or standing. We give the user a goal device in each task, let the user make a selection using gaze, and pull the trigger when feedback is obtained. We require the user to reset the gaze direction to the initial direction before starting each task or before starting the selection task.
We recorded the user's eye movement and task information in the study. After the experiment, we interviewed users and asked them how they felt about the three methods. Finally, we asked users to score the three methods from multiple aspects.

\begin{figure}[htbp]
\centering
\begin{minipage}[t]{0.3\textwidth}
\centering
\includegraphics[width=4.5cm]{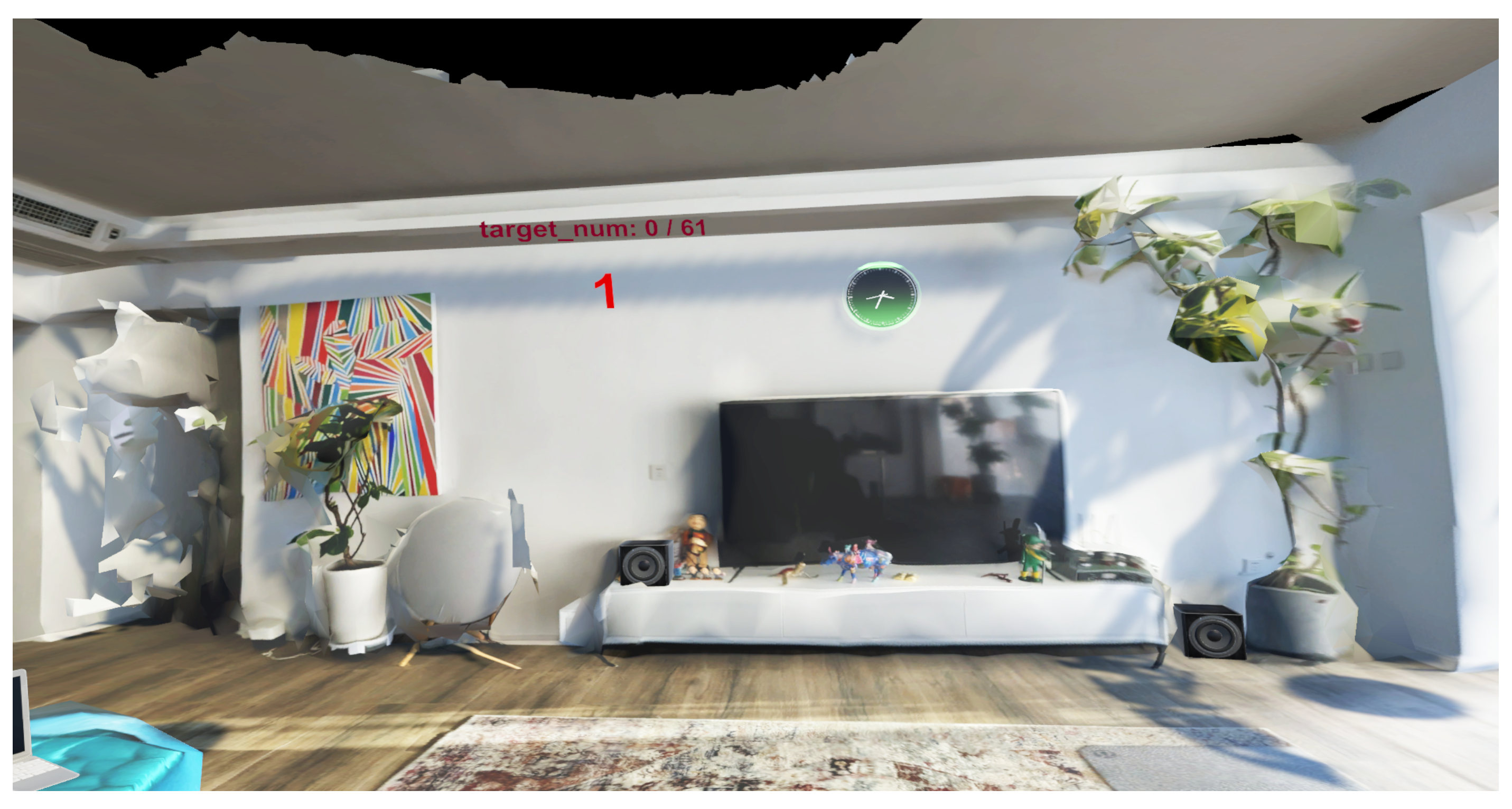}
\label{fig_22}
\end{minipage}
\begin{minipage}[t]{0.3\textwidth}
\centering
\includegraphics[width=4.5cm]{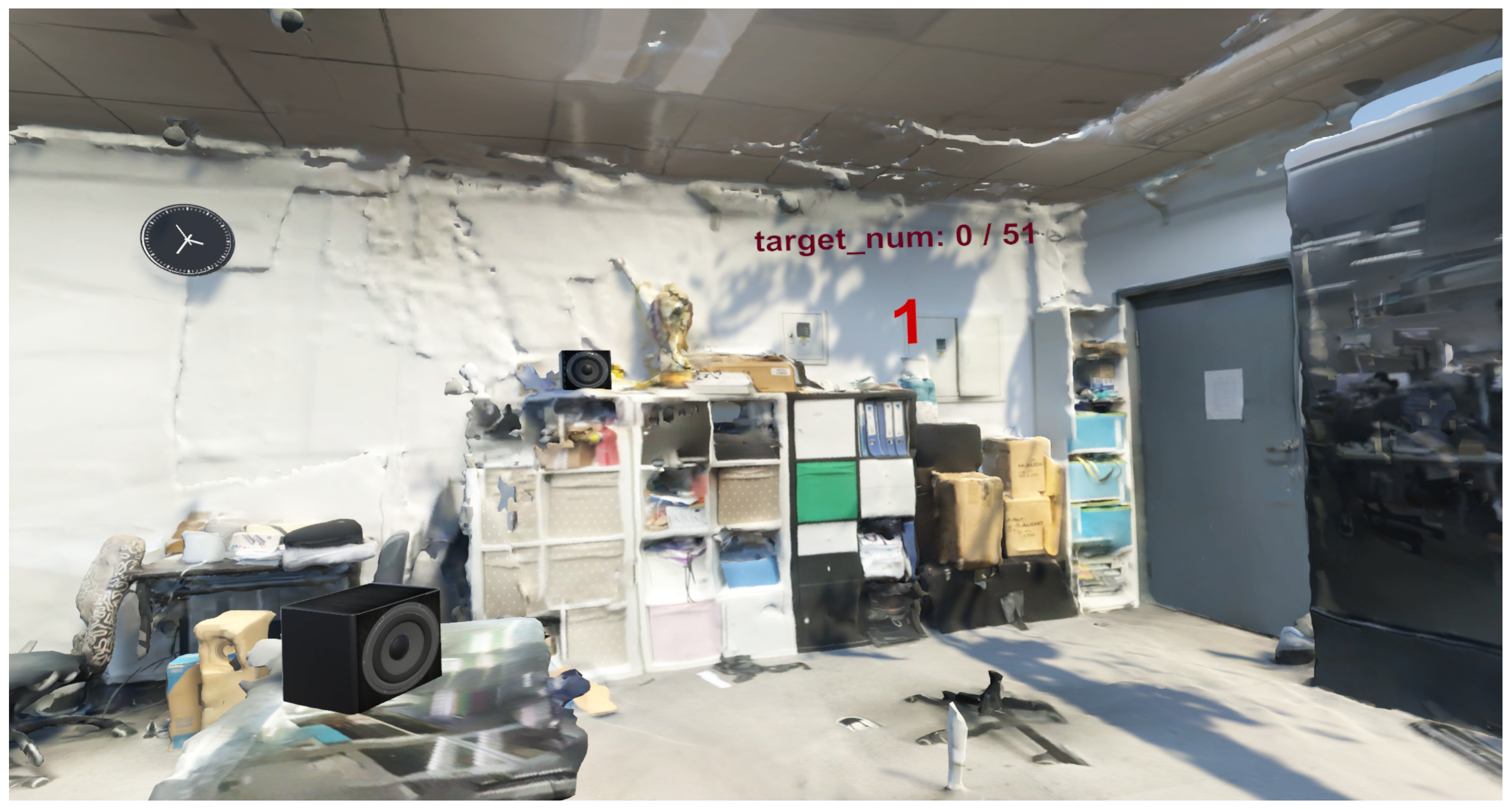}
\label{fig_23}
\end{minipage}
\begin{minipage}[t]{0.3\textwidth}
\centering
\includegraphics[width=4.5cm]{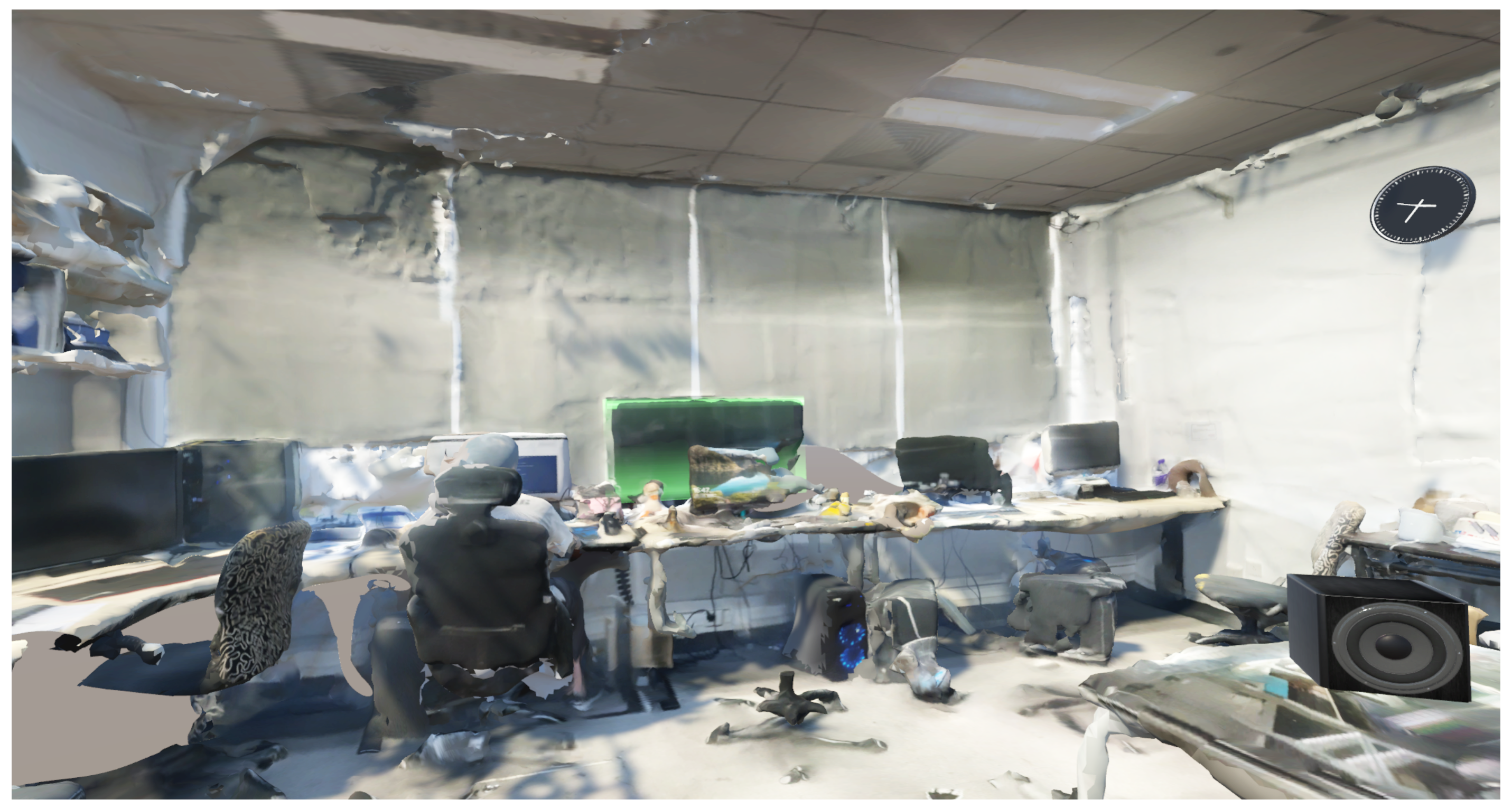}
\label{fig_23}
\end{minipage}
\caption{Scene setting for Study 4. We simulated three daily interaction scenarios (living space(left), working space with more equipment(middle), and working space with less equipment(right)) to test the performance of users using various technologies in different environments.}
\end{figure}

\subsection{Task And Procedure}
Participants were asked to wear a virtual reality headset and calibrate the gaze tracker in the default calibration application of VIVE Pro Eye. An experimenter first explained the experiment to the participants. Then, the experimenter introduced the experiment scene to the user and informed the user of the name and location of each device. Participants spent some time familiarizing themselves with the scene's setting until they remembered the name and location of each device, which is consistent with Study 3. Before users test each method, we will ask users to use the method to try to make selections and then start a formal experiment after the user is familiar with the method. It usually takes 10 minutes for users to familiarize themselves with scenarios and methods.

For each scenario and each technology, each participant performs the equipment number $\times$ 5 target-suggested trials (3 technology $\times$ (12 + 20 + 10) equipment $\times$ 5 times). At the beginning of each trial, the user resets his body and eyes to the starting position. After a three-second countdown, the target name and location will appear in the center of the user's gaze. Then the user will move the head and eyes to indicate the target selection, and when the user is sure that they get the highlight feedback of the target, they will trigger the button on the controller. We require users to complete the interactive task with the target as naturally, quickly, and accurately as possible during the experiment. After completing a scene, the user can rest for 10 minutes and complete the questionnaire. The average duration of the experiment is 50 minutes, with two rests. Finally, we interviewed users and collected their opinions on different technologies.

\subsection{Results}
A total of 7649 valid data were collected in the experiment.
We recorded when the task started, when the system captured the participant's intention, and when the participant confirmed the selection by pressing the trigger. 
Then we calculated four indicators of selection efficiency: 
\begin{itemize}
\item Intent Detect time (DT) (from the start of the task to when the participant's intent was detected)
\item Early Capture time (CT) (from the time the system captures the user's selection target to the user's confirmation)
\item Target Selection time (ST, ST = DT + CT) (from the start of the task to the participant's confirmation of selection)
\item The Number of Errors (E-Num, the number of times that the target estimated by the system is inconsistent with the target selected by the user ).
\end{itemize}
We ran RM-ANOVA tests ($p < 0.05$) on the CT and the ST with the variables of technique and scene, with the post-hoc T-tests ($p < 0.05$).
We highlight our significant results and findings below. 

\textbf{1. \projectName{} achieved higher interaction speed and greater ability to capture intent ahead of time.} 
The baseline requires the user to look at the device accurately, so the speed is relatively poor. For scenarios with many denser devices, users need to control the gaze more accurately using the baseline method, which takes much time. The baseline method only considers that the user has a choice intention when their gaze achieves the goal, so the CT is shorter.
Both projectName{} and KNN significantly reduce the selection time and capture user intent earlier. Compared with KNN, \projectName{} has significantly improved efficiency because the model constructed by projectName{} is more in line with the user's psychological expectations. This result means that the user can quickly complete the target selection when using \projectName{}, which improves the efficiency of device selection, and the system can capture the user's selection intention faster and respond in advance. In addition to the overall efficiency improvement, projectName{} has better interactive performance in special cases.

\begin{figure}
  \includegraphics[width=0.7\textwidth]{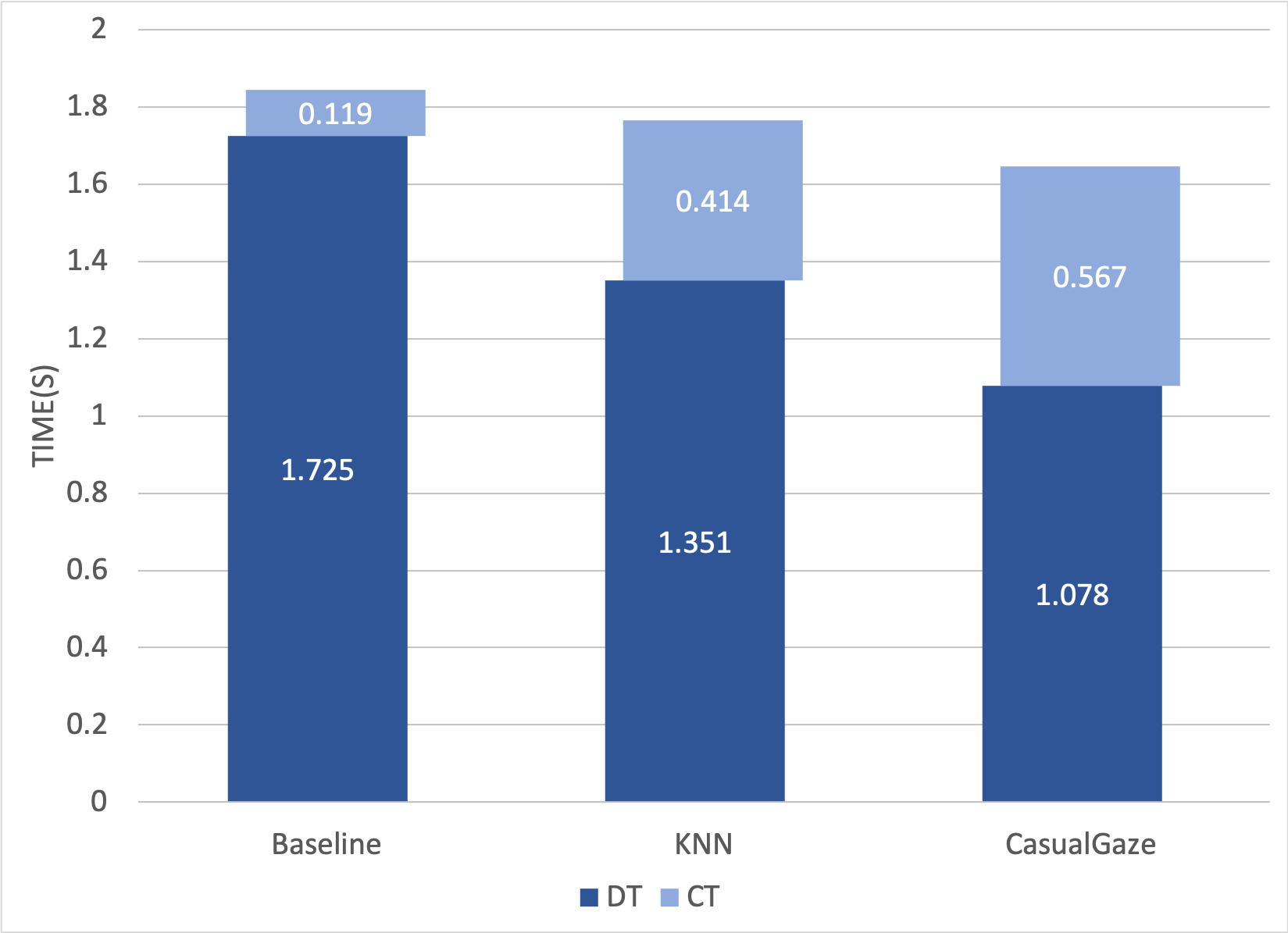}
  \caption{Selection speed and capture speed of different technologies. The top part represents the CT capture the user's intention, and the down part represents the DT for the user to select a target. \projectName{} is significantly better than the other two methods in overall speed and system capture speed.}
  \label{fig:set}
\end{figure}

\textbf{2. \projectName{} has a more significant improvement in the selection of devices with small size.}
The baseline technique for device selection is related to the device size in the user's field of vision. For small devices, it is more difficult for users to keep their gazes on the device, which increases the difficulty of selecting small devices and affects the device selection time. So we compared the selection efficiency of the three techniques for devices of different sizes. Since the distance between the devices and the user is different, we adjusted the size of each device with a standard distance of 3m according to the user-centered spherical coordinate system and then divided the devices into three levels: small devices (diameter less than 0.4m), medium devices (diameter between 0.4m- 0.8m), large equipment (diameter greater than 0.8m).

The RM-ANOVA test showed that there is a significant difference in the speed of device selection for different sizes($ST{mean}:F_{(2,24)} = 3.579, ~p=0.044<0.05$). When using size as the independent variable, since some medium devices are located behind the user, the selection speed of these devices is significantly slower, so we only consider the 32 devices in front of the user. As the size of devices increases, the selection time also decreases, which means the larger the device is, the faster the user makes the selection. For KNN and CasualGlance, the average selection time of users for small devices is shorter than that for medium-sized devices. When using these two methods, the device's size does not pressure the user's intention expression, and their interaction efficiency is improved. For small devices, the average selection time of CasualGlance is 20\% faster than the baseline.

The RM-ANOVA test showed that there are significant differences in the selected speed of different methods($ST{mean}:F_{(2,24)} = 6.182, ~p = 0.007 <0.05$). As shown in the figure, both  \projectName{} and KNN have improved the selection speed of devices of different sizes, and the smaller the device size, the more significant the improvement. When the size of the device is small, the user must accurately look at the device when using the baseline method, which increases the user's gaze control burden and psychological burden, while KNN and  \projectName{} increase the selection threshold of small devices and improve the user's selection efficiency. For large-sized devices, the user's gaze is more likely to fall inside the device, and KNN and  \projectName{} can also optimize the selection threshold of the device, which also shows that when the user makes a natural selection, the gaze does not necessarily fall inside the device. The RM-ANOVA test showed that different techniques also have significant differences in CT for different size devices($CT{mean}:F_{(2,24)} = 31.066, ~p <0.05$). For small devices, the CT of the user obtained by the model is longer, which means that the selection intention of the small device can be captured more quickly, and the user's pressure when selecting a tiny target is reduced.

We also analyzed users' average error numbers when they selected devices of different sizes. Based on user reports, we removed 1\% of mission errors (users chose the wrong target). The RM-ANOVA test showed that both device size and technology have a significant impact on the number of errors in device selection($\text{mean}_{size}:F_{(2,24)} = 47.444, ~p <0.01, \text{mean}_{tech}:F_{(2,24)} = 220.702, ~p <0.01$). For small devices, the number of errors in the baseline is even more than half (2.66 out of 5). It is difficult for users to control the gaze to fall on the device accurately, and KNN and model effectively reduce these errors. The model has the fastest selection speed and the lowest number of errors for devices of different sizes, effectively improving the user's selection efficiency.
\begin{figure}
  \includegraphics[width=0.8\textwidth]{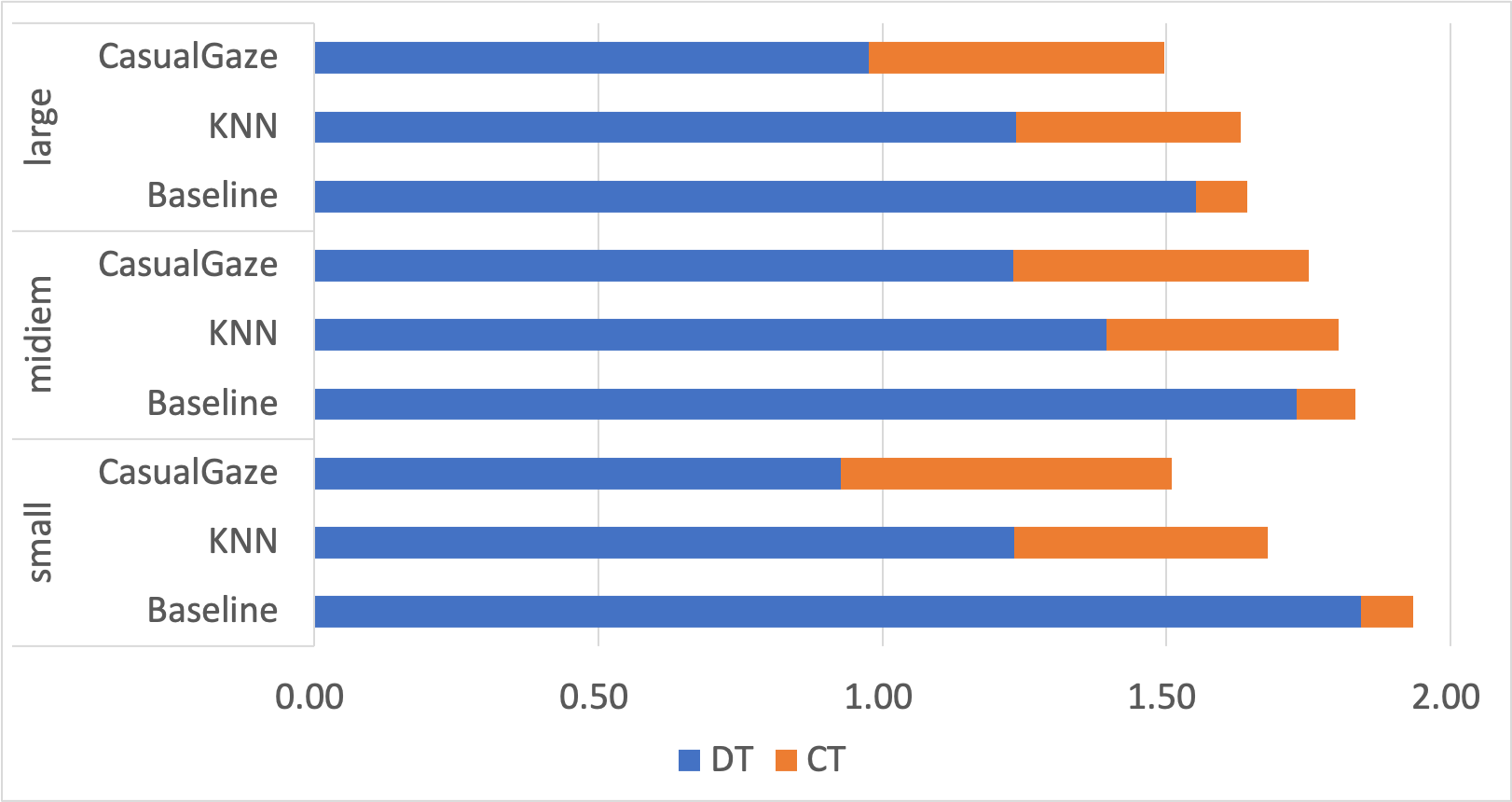}
  \caption{Selection time of different technologies for different size targets.The yellow part represents the CT capture the user's intention, and the blue part represents the DT for the user to select a target. \projectName{} has faster selection and intent capture in different situations.}
  \label{fig:set}
\end{figure}

\textbf{3. \projectName{} has different interaction boosts in different situations.} 
We divided the devices into five different types (N, S, L, C, D) in Study2. To investigate the performance of different techniques on different types, we adjusted the relative distances and sizes of 42 devices in three scenarios in a user-centered spherical coordinate system and classified them.

The results of the RM-ANOVA test show that the device type ($ST{mean}:F_{(4,48)} = 19.624, ~p <0.05$) significantly affects the user's choice time. Different device types reflect different parameters and layouts, and the results show that these variables can significantly affect user behavior. Users will be affected by the target and interfered with by other targets when selecting. Taking the baseline as an example, users often need to spend extra attention to control eye movements or turn their bodies when selecting small targets and special positions, thus increasing the target selection time. For devices close to each other, the user needs to avoid his gaze from falling on the interference target, which increases the user's selection time. The overall time is less for extreme relative size devices due to the choice of large devices. When we only focus on the smaller devices, the selection time will be longer than normal devices, indicating that the interference of other devices will increase the behavior and psychological burden of user selection. \projectName{} reduces the extra burden on users by modeling the natural behavior of users and dramatically improves the selection efficiency of users.

The results of the RM-ANOVA test show that technology ($ST{mean}:F_{(2,24)} = 10.923, ~p <0.05$) significantly affects the user's choice time.
We discussed the performance of the three technologies at different device sizes in the previous conclusions, so we will focus on the last three cases. For devices in special locations, the time for all three techniques increases significantly because the user needs to turn their body. In this case, most of the user's selection time comes from body movement and control, and \projectName{} simplifies the model of specific locations, so there is no noticeable efficient improvement. For devices close to each other, the baseline requires users to spend extra effort to avoid their gaze falling on the interfering device, so it takes longer selection time. KNN and \projectName{} effectively reduce the selection pressure of users, and \projectName{} shortens the selection time by 5\% compared with KNN. For devices with considerable disparities in size, when users choose a large device, their gaze always falls on the device, so we mainly focus on the performance of smaller devices. Compared with the baseline, KNN has a specific improvement, and \projectName{} has a more significant efficiency improvement in this case, which shortens the selection time by 10\% compared to KNN (15\% for the baseline). This result is consistent with the performance of \projectName{} analyzed in Study 3.

\begin{figure}
  \includegraphics[width=0.8\textwidth]{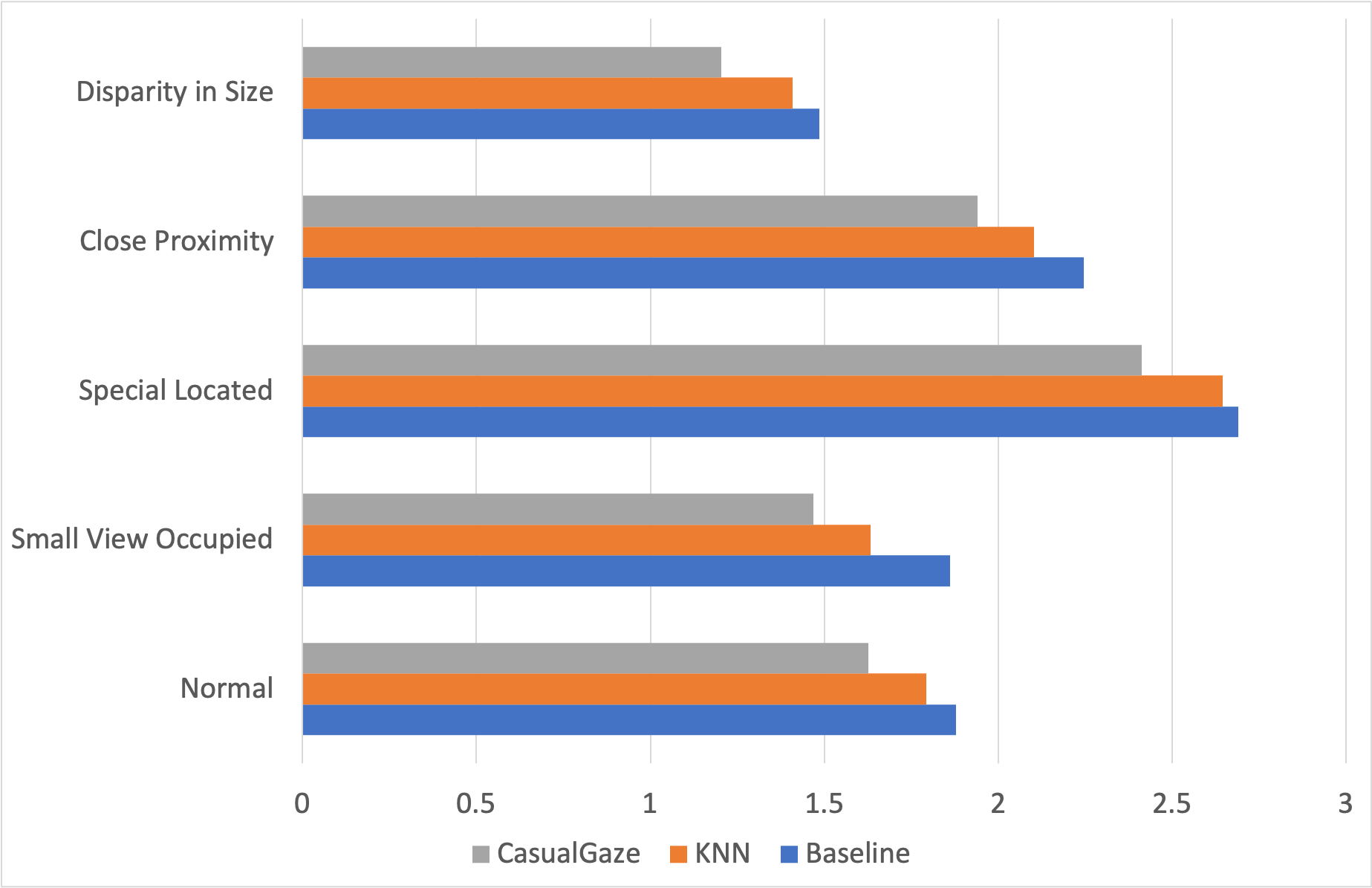}
  \caption{Selection time of different technologies for different cases. \projectName{} also has faster selection and intent capture in different situations.}
  \label{fig:set}
\end{figure}

\textbf{4. Different environments have an impact on the speed of user interaction.}  
RM-ANOVA test shows that different environment settings have significant differences in target selection time and early capture time($\text{mean}:F_{(2,24)} = 26.611, ~p <0.05, \text{mean}:F_{(2,24)} = 5.471, ~p = 0.011<0.05$). Our analysis found that differences in the location and number of devices and user familiarity with the environment have led to differences in user behavior in different environments. The user's selection speed in the living space is significantly faster than in the workspace, and the user's selection speed is the slowest in the workspace with more devices.
In the working space with more devices, because the equipment is placed more densely, users will slow down their interaction speed in order to avoid false touches. When the user is unfamiliar with the environment, the line of sight when selecting the device is more accurate (the data in Study 3 also illustrates this), and the overall selection speed will also decrease. In different environments, we do not require the user's posture. The user can complete the experiment sitting or standing or performing small-scale activities. The data shows that the user's posture has no effect on \projectName{}, and the selected speed of \projectName{} is significantly improved in all three environments, which also shows that \projectName{} has better versatility.


\textbf{5. \projectName{}and KNN reduced requirements for control accuracy while \projectName{} was more stable.} 
At the same time, we recorded the user's subjective evaluation of the three technologies in terms of efficiency, speed, accuracy, fatigue, ease of use, and usability. The specific scores of the subjective scale are shown in the figure.

\begin{figure}
  \includegraphics[width=0.8\textwidth]{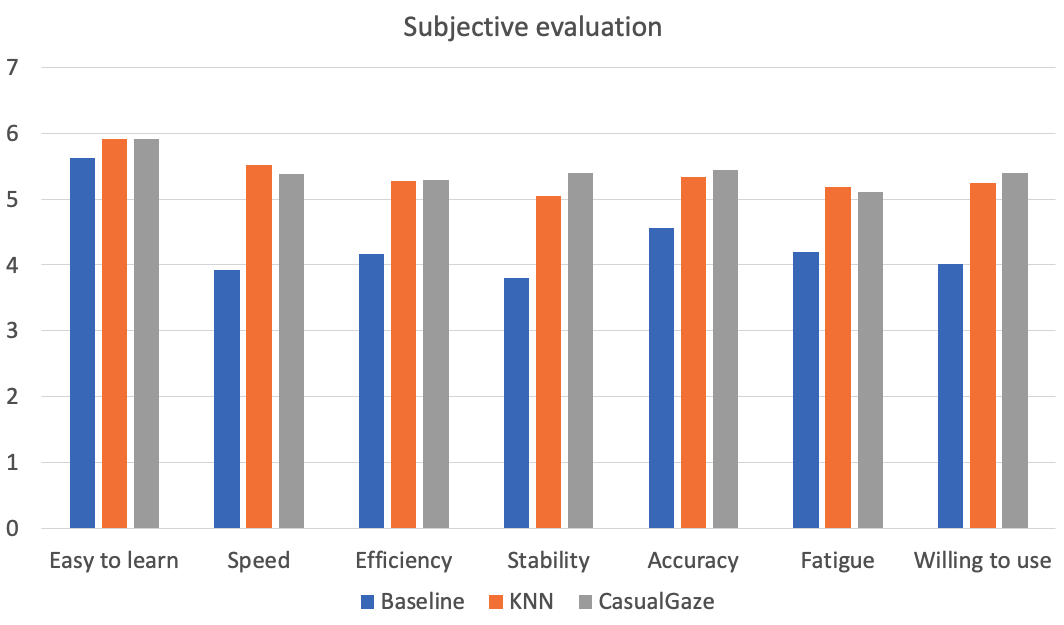}
  \caption{User's subjective ratings for different methods. The scores of the baseline method are generally low. Users rated KNN and \projectName{} in terms of subjective performance such as speed, fatigue, accuracy, Etc. is relatively close, but users believe that the stability of CasualGaze is significantly better than KNN.}
  \label{fig:set}
\end{figure}

Since the three techniques express the selection intention through the user's behavior of looking at the target, the learning cost is low.
Among them, different technologies have significant differences in stability and accuracy. Due to the error of the detective device itself and the difference in the user's control ability, it is difficult for users to ensure that they keep their gaze on the device they want to select for a while. It is even more difficult for devices with small sizes. Both \projectName{} and KNN improve the selection range of the device so that the user can be more relaxed when selecting the device, and the selection of target will be stable. In terms of accuracy, the behavior provided by \projectName{} is more in line with the user's gaze distribution, so it is more effective for distinguishing between multiple devices.
Compared with the other two technologies, \projectName{} does not significantly improve accuracy and speed, but users have unanimously recognized its interactive performance. Users can feel that \projectName{} is more efficient and convenient during use.

We collected some user comments as follows:
\begin{itemize}
    \item User 1: \projectName{} can quickly capture the target I want to select compared with the baseline.
    \item User 2: The selection speed of RNN and \projectName{} seems to be the same, but I think \projectName{} is more stable.
    \item User 3: When I use \projectName{} to switch targets, I will not have as many accidental touches as the other two methods.
    \item User 4: I do not need to be very precise when I use \projectName{} to select a target, which is very convenient for me.
\item User 5: Although the other two techniques can help me with the selection task, after using \projectName{}, I would not like to use the other two techniques.
\end{itemize}

All users expressed that they are willing to use \projectName{} to interact with multiple devices in the smart scene. Users think that \projectName{} can help them more easily and quickly determine the target device they want to choose. At the same time, many users say that they do not notice the difference between \projectName{} and KNN in terms of speed, but think that \projectName{}'s stability was better than that of KNN.

\section{Discussion}
\projectName{} aims to recognize the target objects that users intend to interact with based on their casual gaze behaviors. Throughout two user studies, we modeled the spatial distribution of casual gaze behavior for independent objects and interacting
object pairs as well as users’ casual gaze behavior in an simulated real-world environment, based on which we built \projectName{} recognization algorithms. In offline and real-time evaluation, we showed that \projectName{} provided high recognition accuracy for the device of interest (96\%) and outperformed the baseline method in interaction efficiency and user experience. Based on the results, we discuss the implementation implication and beneficial application of \projectName{}.

\subsection{Influence of More Than One Nearby Devices}
Study 1 tested how a nearby device influences or squeezes the space belonging to the wanted device in users' minds. However, in a more realistic environment, there is more than one nearby device around each device. In the current implementation of \projectName{}, we independently calculated the influence of every nearby device case by case. Although this strategy was proved valid in the offline and real-time evaluation, the exact mechanism of influence of more than one nearby device may differ from simply adding the influences linearly. We did not consider the occlusion case when studying the device effect. When there is occlusion between devices, there may be more differences in user behavior, such as moving the body to express the intent of selecting an occluded object. Characterizing this mechanism and modeling occlusion situation may be a potential way to improve the proposed algorithm's recognition accuracy, and we consider this as future work of great value.

\subsection{Target Attributes Beyond Location and Size}
Study 1 found that the device's location affects its distribution, and we simplified this problem when constructing the model. When a person faces a specific direction, the devices on the side and the back will have a larger distribution radius, which means that for some devices that are difficult to observe, users are more inclined to express the tendency to glance. In Study 1\&2, we abstracted the target device as a sphere to simplify the influence of the device shape. In subsequent experiments, we verified that the model built without considering the device shape is also effective. The appearance of the device affects the distribution of the gaze. For a floor lamp with a larger vertical size, the deviation of the user's line of sight in the two directions will be different from that of a spherical lamp of the same size. Incorporating the device's location and appearance into the model is the direction of our future research.

\subsection{Application Scenarios that Benefit from Attention Awareness}
\projectName{} aims to solve a fundamental research question: how to recognize the attention of users based on casual gaze behavior when they are intended to interact with smart devices. Obtaining the attention points can facilitate the subsequent interaction with the devices. For example, recognizing the interaction intention can be leveraged to save users' efforts in waking up the devices (e.g., by speaking a wake word). Features of the smart environment can also adapt to whether the user is focusing on the devices. For example, the strength of light can be raised when users look at the clock is detected to help read the time.

\section{Conclusion}
In this work, we present \projectName{}, an eye-gaze-based target selection technique to enable casual gaze input, with which the user merely needs to move their gaze towards the target object in a casual object instead of actively staring on it. 
To model casual gaze behavior, we studied the influence of the independent device's parameters (size, location) and the influence caused by the relative distance, position, and size between two nearby devices on the distribution. Further, we collected casual gaze data in a simulated real-world environment to investigate temporal features including speed, randomness features, and patterns in
"blurred areas" for casual gaze. Based on the understanding, we develop \projectName{} algorithms combining spatial Gaussian modeling and temporal optimizations for casual gaze target prediction. Finally, we evaluated \projectName{} and two baseline methods in three environments with different object layouts. Results showed that \projectName{} has higher interaction efficiency, while users agreed that \projectName{} is stable, efficient, and accurate. We believe our work would facilitate future research towards natural gaze-based human interfaces.


\bibliographystyle{ACM-Reference-Format}
\bibliography{sample-base}


\end{document}